\documentclass[final]{IEEEtran}
\pdfoutput=1
\makeatletter
 \def\@textbottom{\vskip \z@ \@plus 1pt}
 \let\@texttop\relax
 
\makeatother
\usepackage[caption=false,font=footnotesize]{subfig}
\usepackage{fancyhdr}
\usepackage{graphicx}
\usepackage[utf8]{inputenc}
\usepackage{amsmath}
\usepackage{amsfonts}
\usepackage{pdfpages}
\usepackage{float}
\usepackage{amssymb}
\usepackage{amsthm}
\usepackage{algorithm}
\usepackage{algorithmic}

\usepackage{cite}
\usepackage{epsf,bm}
\usepackage{epstopdf}

\newtheorem{proposition}{Proposition}
\usepackage{mathtools}
\usepackage{changepage}

\IEEEoverridecommandlockouts
\makeatletter\usepackage{xcolor}
\usepackage{color}
\def\BSTATE{\STATE\hskip-\ALG@thistlm}
\makeatother

\allowdisplaybreaks 

\usepackage{setspace}	

\begin{document}

\title{Stable Throughput of Cooperative Cognitive Networks with Energy Harvesting: Finite Relay Buffer and Finite Battery Capacity}
\author{\large Mohamed A. Abd-Elmagid$^*$, Tamer ElBatt$^\dagger$$^\bullet$, Karim G. Seddik$^\ddagger$, and Ozgur Ercetin$^\diamond$ \\ [.05in]
\small  \begin{tabular}{c} $^*$Wireless@VT, Department of ECE, Virginia Tech, Blacksburg, VA.\\$^\dagger$Computer Science and Engineering Department, American 
University in Cairo, AUC Avenue, Egypt.\\
 $^\bullet$Department of EECE, Faculty of Engineering, Cairo University, Giza, Egypt. \\
$^\ddagger$Electronics and Communications Engineering Department, American University in Cairo, AUC Avenue, Egypt.\\
$^\diamond$Faculty of Engineering and Natural Sciences, Sabanci University, Istanbul, Turkey.\\

email: maelaziz@vt.edu, telbatt@ieee.org, kseddik@aucegypt.edu, oercetin@sabanciuniv.edu
\end{tabular} 

\thanks{This work was supported 
in part by the Egyptian National Telecommunications Regulatory Authority (NTRA), and was done when Mohamed A. Abd-Elmagid and Tamer ElBatt were with WINC, Nile University, Egypt. In addition, it was presented in part at the IEEE International Conference on Computing, Networking and Communications (ICNC), 2017 \cite{Conference}.}
}

\maketitle
\begin{abstract}
This paper studies a generic model for cooperative cognitive radio networks where the secondary user is equipped with a finite relay queue as well as a finite battery queue. Our prime objective is to characterize the stable throughput region. Nevertheless, the complete characterization of the stable throughput region for such system is notoriously difficult, since the computation of the steady state distribution of the two-dimensional Markov Chain (MC) model for both finite queues is prohibitively complex. We first propose an algorithm to characterize the stable throughput region numerically, and show its sheer computational complexity for large queue lengths. To lend tractability and explore the nature of design parameters optimization at the cognitive node, we next focus on two simpler systems, namely, finite battery queue with infinite relay queue and finite relay queue with infinite battery queue (referred henceforth as dominant system $1$ and $2$, respectively). For each proposed dominant system, we investigate the maximum service rate of the cognitive node subject to stability conditions. Despite the complexity of the formulated optimization problems, due to their non-convexity, we exploit the problems' structure to transform them into linear programs. Thus, we are able to solve them efficiently using standard linear programming solvers. Our numerical results demonstrate that, in practical systems, finite battery and relay queues achieve the same level of benefits of a system with infinite queue sizes, when their sizes are sufficiently large. They also reveal that the achievable stable throughput region significantly expands when the arrival rate of the energy harvesting process increases.

\end{abstract}

\IEEEpeerreviewmaketitle
\section{Introduction}

One of the prominent challenges in wireless communication networks is to efficiently utilize the spectrum. The cognitive radio technology has the potential to improve the utilization of the scarce spectrum resource. In cognitive radio networks, better utilization of spectrum is made possible by allowing unlicensed (i.e., 
secondary) users (SUs) access the spectrum owned by the licensed (i.e., primary) users (PUs) \cite{1,2,liang2011cognitive} using spectrum underlay or spectrum overlay access techniques. Unlicensed SUs sense the spectrum for activity of licensed PUs \cite{sensing1,sensing2}, and based on the sensing information, spectrum access decisions are made by the SUs. In the spectrum underlay paradigm \cite{underlay1,underlay2}, SUs transmit even if PUs are sensed to be present. Nevertheless, the spectrum occupation of the SUs is tied with a minimum quality of service (QoS) guaranteed for the PUs. This, in turn, calls for the necessity of designing efficient spectrum access schemes to maximize the SU's achievable throughput while satisfying the PU's QoS constraints. In the spectrum overlay paradigm \cite{overlay1,overlay2}, SUs only access the spectrum when PUs are sensed to be idle. However, due to spectrum sensing errors, collisions are experienced by both PUs and SUs as a result of the interference between their transmissions.

Energy harvesting has recently emerged as a promising technology to prolong the life time of energy-constrained wireless networks. This is triggered by the fact that energy harvesting circuitries provide wireless devices with the capability of perpetual charging of their batteries via harvesting energy from the surrounding environment. Significant research has been conducted on wireless powered communication networks from different perspectives and with the focus on different performance aspects \cite{UluYenJ2015,Zhang_throughput_maximization,Abd-Elmagid2015,77,100,Abd-Elmagid2017,6552840,7996351,7332956,Abd-Elmagid2016}. Incorporating energy harvesting capability to cognitive radio networks has attracted considerable attention in the literature~\cite{R1_baseline,R1,R2,R3,R4,144,155}. In \cite{R1_baseline}, the authors studied a non-cooperative cognitive radio network composed of two primary and secondary source-destination pairs of nodes. The primary source node (PS) is equipped with an energy queue and is assumed to be solely powered by energy harvesting. The secondary source node (SS) is not only able to access the channel when the PS is idle, but also is allowed to transmit data with probability $p$ whenever the PS is active. The goal was to characterize the optimal transmission probability of the SS, $p^*$, that maximizes its achievable throughput while maintaining the stability of the primary source packet queue at given packet arrival and energy harvesting rates. \cite{R1} extended the analysis of \cite{R1_baseline} and characterized $p^*$ for the two other potential scenarios: i) SS is solely powered by energy harvesting whereas PS is plugged to a reliable power supply, and ii) both PS and SS have energy harvesting capabilities. Differently from~\cite{R1_baseline,R1}, the authors in \cite{R2} assumed that the SS can sense the channel perfectly and may only access the channel if the PS is idle. Whenever the SS has the opportunity to access the channel, it can consume $i$ energy packets with probability $p_i$ such that $i$ is less than or equal to the total number of energy packets in its energy queue. The objective was to optimally tune $p_i$ so as to maximize the achievable throughput of the SS. In \cite{144}, the authors investigated the optimal spectrum sensing policy to maximize the expected total throughput subject to two constraints, namely, an energy causality constraint and a collision constraint. The objective of the energy causality constraint is to guarantee that the total consumed energy at the cognitive node is less than or equal to the total harvested energy, while the collision constraint protects the primary user by guaranteeing a minimum QoS requirement. The optimal transmission power and density, for the cognitive nodes, were characterized in \cite{155} so that the secondary network throughput is maximized under given outage probability constraints in, both, the primary and secondary networks.

 Cooperative cognitive radio networks (CCRNs) have recently attracted considerable attention \cite{R3,R4,simeone2007stable,sadek2007cognitive,kompella2011stable,3,4,kulkarni2016stable,5,hyder2016interference}. The notion of cooperation in cognitive radio networks is that the SU helps the PU in successfully transmitting its data packets to the destination so as to decrease the number of time slots dedicated to retransmit the PU's lost packets over its direct link. This, in turn, enhances the available time slots for the SU to access the channel and transmit its own data packets. Therefore, both the PU and SU benefit from the cooperation. The most relevant literature can be categorized into two sets: i) CCRNs with energy harvesting capability~\cite{R3,R4}, and ii) CCRNs without energy harvesting capability~\cite{simeone2007stable,sadek2007cognitive,kompella2011stable,3,4,kulkarni2016stable,5,hyder2016interference}. Incorporating a relay queue at the SS to the system setup in \cite{R1}, \cite{R3} formulated the maximum weighted sum of the service rates at the SS queues problem as a Markov Decision Process (MDP), and \cite{R4} characterized the stable throughput region for Poisson energy harvesting processes using the Dominant System approach~\cite{8}. In \cite{simeone2007stable}, SU acts as a relay for delivering the PU's data packets, wherein it maximizes its achievable throughput for a given fixed throughput value demanded by the PU by optimizing its transmit power. \cite{sadek2007cognitive} proposed two multiple-access protocols in a cooperative cognitive radio network consisting of $M$ source terminals, a single cognitive relay and a single common destination. In \cite{kompella2011stable}, the SU was allowed to share the channel with the PU, and could act as a relay for transmitting successfully decoded PU's packets that were not successfully decoded by its destination. For this proposed channel access scheme with fixed scheduling probability, the stable throughput region was characterized. \cite{3} introduced a full cooperation protocol in a wireless multiple-access system for a system composed of $N$ users wherein each user is a source and at the same time a potential relay.

In \cite{4}, the authors proposed a cooperative strategy with probabilistic relaying. In this strategy, the SU is equipped with two infinite length queues; one is for storing its own packets and the other is for relaying the PU packets. If the PU's packet is not successfully decoded by the destination, whereas it is successfully decoded by the SU, the SU admits the PU's packet with probability $a$. On the other hand, when the PU is sensed idle, the SU serves its own data queue with probability $b$ or the relay queue with probability $1 - b$. The authors in \cite{5} characterized the throughput region when the relay buffer at the SU has finite length, and in \cite{kulkarni2016stable} generalized the model studied in \cite{4} to the scenario of having multiple cognitive nodes, and characterized the optimal transmission probability of SU that maximizes the individual achievable throughputs. In \cite{hyder2016interference}, the authors incorporated the overhead of forming the cooperation and the potential interference caused by SUs' transmissions. The system is modeled as a Markov decision process (MDP), and the impact of cooperation overhead and secondary interference are quantified in the actions and their rewards. Note that in \cite{4,kulkarni2016stable,5,hyder2016interference}, it was implicitly assumed that the SU is equipped with unlimited energy supply, i.e., the SU does not suffer from any energy limitations whenever it has the opportunity to access the channel. It is worth noting that a similar system setup to that in \cite{4} has recently studied in the context of cooperative relay networks~\cite{R6,1_Cop}. Incorporating energy harvesting to the considered system model in \cite{4}, the authors in \cite{R6,1_Cop} assumed that both the source and relay nodes are equipped with infinite battery queues to store their harvested energy. Unlike the probabilistic selection strategy (selection of whether the transmitted packet is from the data queue or the relay queue) adopted by the SU whenever it is able to access the channel in \cite{4}, the relay node does not differentiate between whether the transmitted packet belongs to the overheard packets from the source node or its own arrival packets. This is due to the fact that, in contrary to CCRNs, within the context of cooperative relay networks, any transmitted packet from the relay node is counted in its own throughput. 

In this paper, we consider a spectrum overlay cooperative cognitive network with a single PU and a single SU. The SU is solely powered by energy harvesting and its battery is replenished by a stochastic energy harvesting process independent from packet arrivals and scheduling decisions. The SU can sense the channel perfectly and is equipped with two data queues; one queue stores its own data packets whereas the other one (relay queue) stores the the overheard unsuccessfully transmitted packets by the PU. A probabilistic strategy is adopted by the SU in both admitting PU's data packets at its relay queue and selecting one of its data queues whenever it has the chance to access the channel. Unlike prior work, the SU has a finite relay to store the packets overheard from the PU and a finite capacity battery to transmit both its own packets as well as the overheard packets of the PU. Compared to the usual assumption in the literature of considering infinite queue lengths, having two finite queues (relay and battery queues) simultaneously adds another layer of complexity to the performance analysis. It can be contemplated that the proposed system model constitutes an important step towards real systems with all finite queue lengths. Our prime objective is to optimally tune the admission and selection probabilities as a function of the relay and battery queue lengths to maximize the SU's achievable throughput.

The main contributions of this paper are summarized as follows:
\begin{itemize}
\item We introduce a generic model for cooperative cognitive radio networks, where the relaying SU is equipped with a finite capacity battery and a finite relay queue. We exemplify the challenges of characterizing the stable throughput region, by demonstrating the intractability of obtaining a closed-form expression of the steady state distribution of the underlying two-dimensional Markov Chain. We propose a numerical approach to characterize the stable throughput region. However, even the numerical solution becomes intractable for large queue sizes.
\item Motivated by the problem complexity, we introduce two simpler problems for the relaying SU, i.e., finite battery queue with infinite relay queue (dominant system $1$) and finite relay queue with infinite battery queue (dominant system $2$). The stability conditions for the two dominant systems are derived.
\item We formulate two optimization problems to characterize the maximum achievable throughput of the SU, subject to the queue stability conditions, for each of the two dominant systems. Despite the fact that the optimization problems are non-convex, we exploit the problems' structure to re-cast them as linear programs. This, in turn, leads to efficient solutions by standard optimization tools.
\item Our numerical results reveal interesting insights about the effects of finite relay and energy queues as well as the energy limitations on the achievable stable throughput region. Specifically, they quantify: 1) the expansion in the throughput region due to increasing the battery queue size, and 2) the enhancement of the maximum sustainable arrival rate of PU's data packets, corresponding to a non-zero SU's achievable throughput, due to increasing the relay queue size. They also reveal that, in practical systems, finite battery and relay queues of sufficiently large sizes are enough to achieve the same level of performance of a system with infinite queue sizes. Moreover, the results demonstrate the great influence of the arrival rate of the energy harvesting process at the SU on the achievable stable throughput region.
\end{itemize}

The rest of the paper is organized as follows. Section~\ref{sec:sys} describes the system model. The stable throughput region of our generic proposed CCRNs with both finite relay and battery queues is characterized in Section~\ref{sec:first}. In Section~\ref{sec:second}, we provide the stability conditions for dominant system $1$, formulate the stable throughput region optimization problem and show how to solve it. The stability conditions for dominant system $2$, and its associated stable throughput region optimization problem are presented in Section~\ref{sec:third}. We present our numerical results in Section~\ref{sec:num}. Finally, Section~\ref{sec:con} concludes the paper.

\begin{figure}
\centering
\includegraphics[width=0.8\columnwidth]{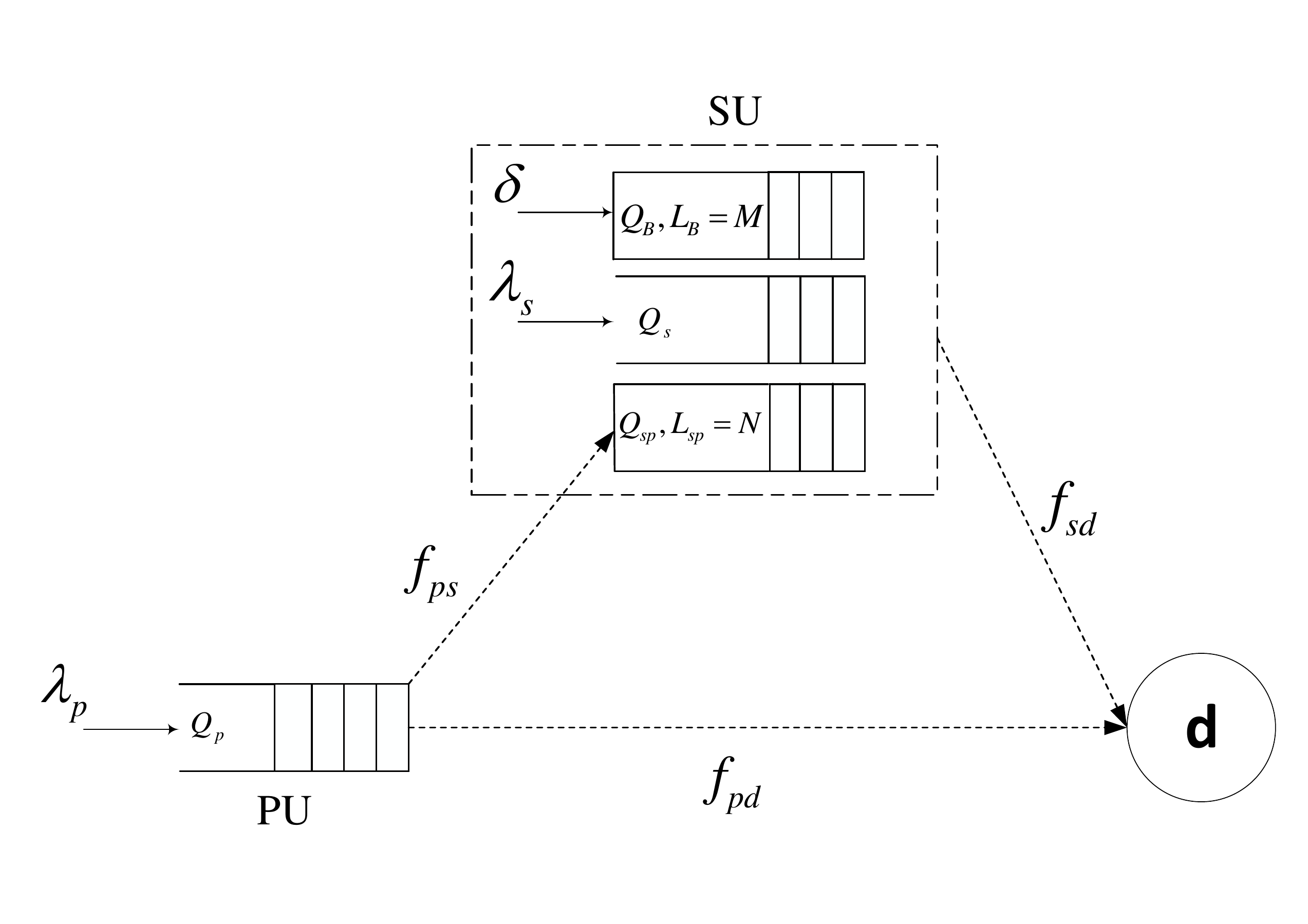}
    \caption{System model.}
     \label{fig:1}
\end{figure}

\section{System Model}
\label{sec:sys}
In this paper, we study a cooperative cognitive radio network as shown in Fig. 1. The network consists of a PU and a SU transmitting their packets to a common destination $d$. The PU is equipped with an infinite queue $(Q_{p})$ for storing its data packets. On the other hand, the SU is equipped with an infinite queue $(Q_{s})$ for storing its data packets and a finite queue $(Q_{sp})$ of length $N$ for storing packets overheard from the PU. The arrival processes at the data queues, $Q_{p}$ and $Q_{s}$, are modeled as Bernoulli processes with means $\lambda_{p}$ and $\lambda_{s}$ \cite{6}, respectively, where $0 \leq \lambda_{p}, \lambda_{s} \leq 1$. The arrival processes at both users are assumed to be independent of each other, and are independent and identically distributed across time slots. SU is equipped with an energy harvesting circuitry to generate energy to perform transmissions. The harvested energy is stored in a finite size battery modeled as a queue $(Q_{B})$ of maximum length $M$. For mathematical tractability, the harvested energy is assumed to be harvested in quantas of size necessary for one transmission attempt. The energy harvesting process\footnote{Note that the energy harvesting process at the SU is assumed to have an arrival rate always sufficient to keep the node alive to listen to the channel at every slot and or transmit ACK packets whenever needed. This can be ensured by designing the energy harvesting system appropriately, e.g., having at least a 
certain size solar panel, or having higher and/or more frequent RF signal transmissions towards the SU. However, given this constant leakage of energy to keep the node alive, we are still short of energy to transmit, and arrivals in addition to leakage rate is assumed to be Bernoulli-like process.} at the SU is modeled as a Bernoulli process with mean $\delta$ \cite{6}, where $0 \leq \delta \leq 1$.

The queue sizes of $Q_p$ and $Q_s$ evolve as follows

\begin{equation}\label{eq1}
Q^{t+1}_{i} = (Q^{t}_{i} - Y^{t}_{i})^{+} + X^{t}_{i}, i \in (p, s),
\end{equation}
where $Q^{t}_{i}$ is the number of packets at the beginning of time slot $t$, and $X^{t}_{i}$ and $Y^{t}_{i}$ are binary random variables that denote the number of arriving and departing packets, respectively. In addition, $(Z)^{+} = \max(Z, 0)$.

 Time is slotted and one slot duration is equal to one packet transmission time. It is assumed that the PU and SU are perfectly synchronized and the SU has perfect spectrum sensing\footnote{Note that imperfect spectrum sensing leads to collisions experienced by both PU and SU due to the possible interference between their transmissions. This, in turn, degrades the achievable throughputs, and, reduces the stable throughput region.} capability. For a successful transmission, the channel should not be in outage, i.e., the received signal-to-noise ratio (SNR) at the destination should not be less than a pre-specified threshold required to successfully decode the received packet. Let $f_{pd}$, $f_{sd}$ and $f_{sp}$ denote the probability of successful transmission between the PU and destination, the SU and destination, and the SU and PU, respectively. We assume that $f_{pd} < f_{sd}$\footnote{Note that when $f_{pd} > f_{sd}$, storing the PU's unsuccessfully transmitted packets to the destination at relay queue is not beneficial to the PU. Specifically, it becomes more efficient for the PU to retransmit any unsuccessfully decoded packet at the destination than transmitting it through the SU's relay queue. This is due to the fact that in such scenario, the PU has a better channel to the destination than that from the SU.} so that we can characterize the relaying role of the SU for the PU. Moreover, acknowledgement packets (ACKs) are sent either by the destination for successfully-decoded packets from the PU or SU, or by the SU for successfully-decoded overheard packets from the PU. In order to obtain analytical characterization, we assume that ACKs are instantaneous, error-free and can be heard by all the nodes in the network similar to \cite{4}.

The proposed channel access policy is as follows. The PU has the priority to transmit a packet whenever $Q_{p}$ is non-empty. If the packet is successfully decoded by the destination, the destination sends back an ACK heard by both users (PU and SU). Therefore, the packet is dropped from $Q_{p}$ and exits the system. If the packet is not successfully decoded by the destination but successfully decoded by the SU, $Q_{sp}$ either admits the packet with probability $a_{i,j}$ or discards it with probability $(1-a_{i,j})$, $i=0,\cdots,M$ and $j=0,\cdots,N$. The packet admission probabilities depend on the number of packets in $Q_{B}$ and $Q_{sp}$, i.e., $a_{i,j}$ is the \textit{admission probability} when $Q_B$ has $i$ packets and $Q_{sp}$ has $j$ packets. This admission strategy, in turn, constitutes the probabilistic admission relaying policy. If the packet is buffered in $Q_{sp}$, the SU sends back an ACK to announce the successful reception of PU's packet. Thus, the packet is dropped from $Q_{p}$ and the SU becomes responsible for delivering the PU's packet to the destination. Finally, if the packet is neither successfully decoded by the destination nor SU, or it is decoded by the SU but not admitted to $Q_{sp}$, then the packet is kept at $Q_{p}$ for retransmission in the subsequent time slot. 

When the PU is idle, the SU's packet transmission depends on the status of battery and data queues. If the battery queue is empty, then the SU is unable to transmit a packet. On the other hand, if the battery queue is not empty, the SU either transmits a packet from $Q_{s}$ with probability $b_{i,j}$ or from $Q_{sp}$ with probability $(1-b_{i,j})$, $i = 0,\cdots,M$ and $j = 0,\cdots,N$. Also, note that the queue selection probability depends on the number of packets in $Q_{B}$ and $Q_{sp}$, i.e., $b_{i,j}$ is the selection probability when $Q_B$ has $i$ packets and $Q_{sp}$ has $j$ packets. If the destination successfully decodes the packet, it sends back an ACK heard by the SU. Therefore, the packet is dropped from its respective queue, i.e., $Q_{s}$ or $Q_{sp}$, and exits the system. Otherwise, the packet is kept at its queue for retransmission. In Section \ref{sec:first}, we characterize the stability conditions when all queues in the network may have infinite size.

\section{Generalized CCRNs with both Finite Battery and Relay Queues}
\label{sec:first}
In this section, we characterize the stable throughput region of the proposed system model. We start by deriving the stability conditions of infinite length queues ($Q_{p}$ and $Q_{s}$). Next, we propose a discrete time two-dimensional Markov Chain (MC) model for finite queues ($Q_B$ and $Q_{sp}$). Then, we show the complexity of characterizing the steady state distribution for the underlying two-dimensional MC with diagonal transitions. Afterwards, we characterize the stable throughput region numerically showing the impact of different system design parameters. Finally, we propose a simpler scheme in which the admission and selection probabilities ($a_{i,j}$ and $b_{i,j}$) are the same among most of the states, to accommodate the sheer complexity of numerical computations as the queues' lengths become large.
\subsection{Stability conditions for infinite queues ($Q_{p}$ and $Q_{s}$)}
Loynes theorem \cite{7} provides the stability condition for an infinite size queue. The theorem states that if the queue arrival and service processes are stationary, the queue is stable if and only if the packet arrival rate is strictly less than the packet service rate. Note that $Q_{B}$ and $Q_{sp}$ are finite queues; therefore, the number of packets in each of them will never grow to infinity since it is upper bounded by $M$ and $N$, respectively.

A packet leaves $Q_p$ if it is either successfully decoded by the destination or successfully decoded by the SU and admitted to the relay buffer $(Q_{sp})$. Therefore, the service rate of $Q_{p}$ is given by

\begin{align}\label{eq2}
\mu_p = f_{pd} + \left(1-f_{pd}\right) f_{ps} \sum_{i=0}^{M}\sum_{j=0}^{N}{a_{i,j} \pi_{i,j}},
\end{align}
where $\pi_{i,j}$ denotes the steady state probability that $Q_{B}$ has $i$ packets and $Q_{sp}$ has $j$ packets at a given time slot, $i= 0,\cdots, M$ and $j = 0,\cdots, N$. Therefore, the stability condition for $Q_{p}$ is given by
 \begin{align} \label{eq3}
\lambda_{p} < f_{pd} + \left(1-f_{pd}\right) f_{ps} \sum_{i=0}^{M}\sum_{j=0}^{N}{a_{i,j} \pi_{i,j}}.
\end{align}
 
 Similarly, a packet leaves $Q_{s}$ if $Q_{p}$ is empty with probability $1 - \dfrac{\lambda_{p}}{\mu_{p}}$, $Q_{B}$ is not empty or it is empty but there is an energy packet arrival, $Q_{s}$ is selected for transmission with probability $b_{i,j}$ and the destination successfully decodes the packet with probability $f_{sd}$. Thus, the stability condition for $Q_{s}$ can be expressed as

\begin{align}\label{eq4}
\lambda_{s} < \left(1 - \dfrac{\lambda_{p}}{\mu_{p}}\right) f_{sd} \left(\sum_{i=1}^{M}\sum_{j=0}^{N}{b_{i,j} \pi_{i,j}} + \delta \sum_{j=0}^{N}{b_{0,j} \pi_{0,j}}\right).
\end{align}

Note that the service rates of $Q_{s}$ and $Q_{sp}$ depend on the state of the battery queue $(Q_{B})$ at the secondary user and vice versa. This dependency, in turn, leads to an interacting system of queues and complicates the characterization of the stable throughput region. In the sequel, we use the concept of Dominant System approach \cite{8} where we assume that $Q_{s}$ and $Q_{sp}$ have dummy packets to transmit when they are empty and, hence, the service rate of $Q_B$ becomes only dependent on the PU's status. This system stochastically dominates our system since the lengths of the SU's queues in the dominant system are always larger than that of our system if both systems start from the same initial state, have the same arrivals and encounter the same packet losses.

 Now, we calculate the steady state distribution of $Q_{B}$ and $Q_{sp}$ ($\pi_{i,j}$) in order to fully characterize the stable throughput region. Consequently, we will investigate the optimal admission and selection probabilities ($a_{i,j}$ and $b_{i,j}$) in order to maximize the service rate of SU subject to queue stability conditions.
 
\begin{figure}
\centering
\includegraphics[width=0.75\columnwidth]{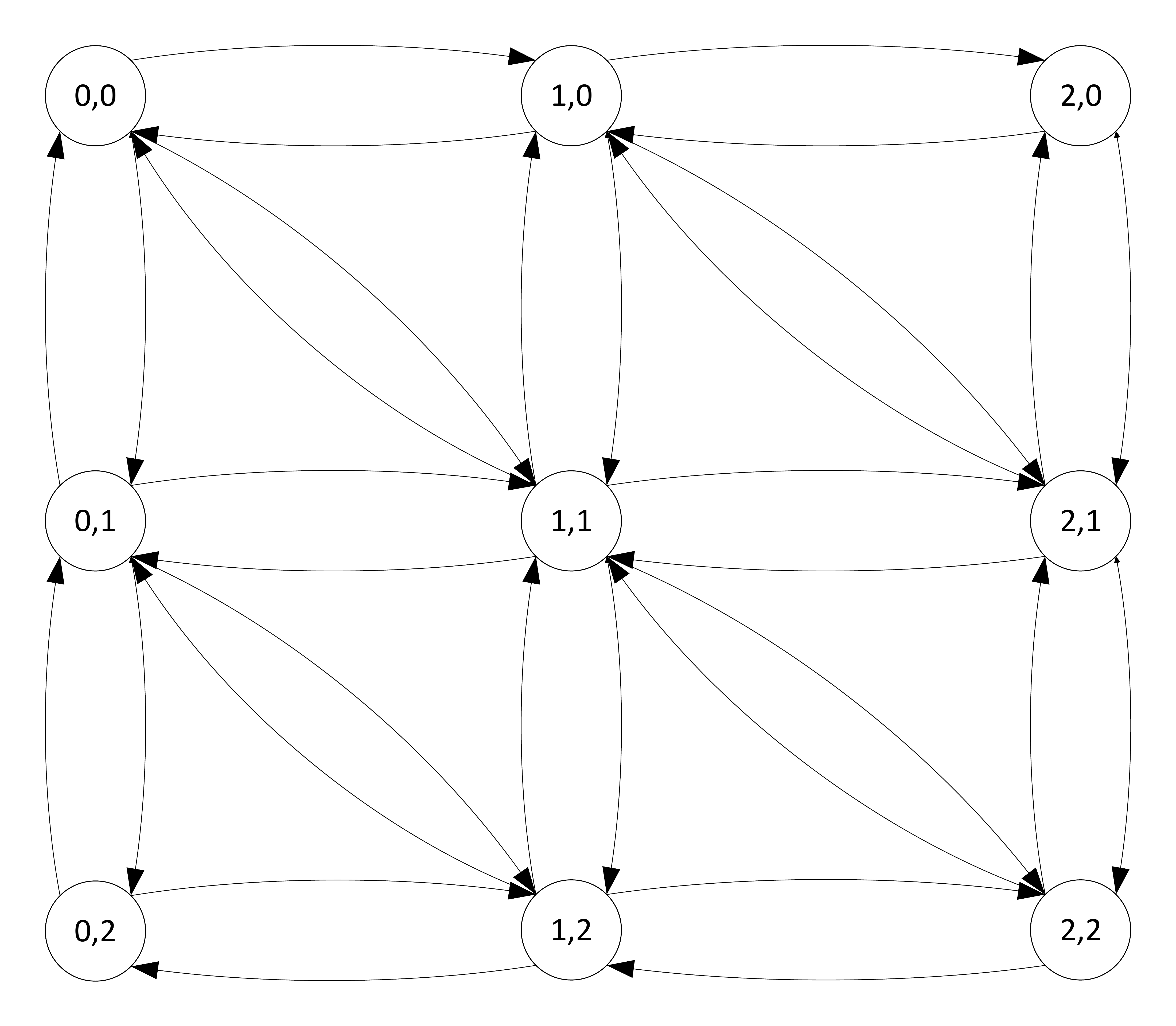}
    \caption{Discrete time two-dimensional MC model for $Q_{sp}$ and $Q_{B}$, where $M = N = 2$.}
     \label{fig:2}
\end{figure}
 \subsection{Discrete time two-dimensional MC model for $Q_B$ and $Q_{sp}$}
 $Q_B$ and $Q_{sp}$ can be modeled as a discrete time two-dimensional MC. The MC is shown in Fig. 2 where state ($i,j$) denotes that the number of packets in $Q_B$ and $Q_{sp}$ are $i$ and $j$, respectively. The probability of moving from state ($i,j$) to state ($i+1,j+1$) is the probability that  $Q_p$ is non-empty, an energy packet arrives at $Q_{B}$, the PU's packet is not successfully decoded at the destination, the SU successfully decodes the packet and $Q_{sp}$ admits the packet. Hence, $P_{i,j \to i+1,j+1}$ can be expressed as
 \begin{align}\label{eq5}
 P_{i,j \to i+1,j+1} = \dfrac{\lambda_{p}}{\mu_p} \delta \left(1 - f_{pd}\right) f_{ps} a_{i,j}.
 \end{align}
  
  The probability of moving from state ($i,j$) to state ($i-1,j-1$) is the probability that $Q_p$ is empty, there is no energy packet arrival at $Q_{B}$, $Q_{sp}$ is selected for transmission and the packet is successfully decoded at the destination. Therefore, $P_{i,j \to i-1,j-1}$ is given by
 \begin{align}\label{eq6}
 P_{i,j \to i-1,j-1} = \left(1 - \dfrac{\lambda_{p}}{\mu_{p}}\right) \left(1 - \delta\right) f_{sd} \left(1 - b_{i,j}\right).
 \end{align}
 
  The probability of moving from state ($i,j$) to state ($i+1,j$) is the probability that $Q_p$ is non-empty, there is an energy packet arrival at $Q_{B}$ and the PU's packet is either successfully decoded by the destination or not successfully decoded by the destination but not successfully decoded at the SU and admitted at the same time. Therefore, $P_{i,j \to i+1,j}$ is given by
 \begin{align}\label{eq7}
 P_{i,j \to i+1,j} = \dfrac{\lambda_{p}}{\mu_{p}} \delta \left( f_{pd} + \left( 1 - f_{pd}\right) \left( 1 - f_{ps} a_{i,j}\right)\right),
 \end{align}
 
  The probability of moving from state ($i,j$) to state ($i,j+1$) is the probability that $Q_p$ is non-empty, there is no energy packet arrival at $Q_{B}$ and the PU's packet is not successfully decoded by the destination but successfully decoded by the SU and admitted to the relay buffer. Therefore, $P_{i,j \to i,j+1}$ is given by
 \begin{align}\label{eq8}
 P_{i,j \to i,j+1} = \dfrac{\lambda_{p}}{\mu_{p}} \left( 1 - \delta\right) \left( 1 - f_{pd}\right) f_{ps} a_{i,j},
 \end{align}
 
  The probability of moving from state ($i,j$) to state ($i,j-1$) is the probability that $Q_p$ is empty, there is an energy packet arrival at $Q_{B}$, $Q_{sp}$ is selected for transmission and the packet is successfully decoded at the destination. Therefore, $P_{i,j \to i,j-1}$ is given by
  \begin{align}\label{eq9}
 P_{i,j \to i,j-1} = \left(1 - \dfrac{\lambda_{p}}{\mu_{p}}\right) \delta f_{sd} \left( 1 - b_{i,j}\right),
 \end{align}
 
  The probability of moving from state ($i,j$) to state ($i-1,j$) is the probability that $Q_p$ is empty, there is no energy packet arrival at $Q_{B}$ and $Q_{s}$ is selected for transmission or $Q_{sp}$ is selected for transmission but the transmitted packet is not successfully decoded by the destination. Therefore, $P_{i,j \to i-1,j}$ is given by
  \begin{align}\label{eq10}
 P_{i,j \to i-1,j} = \left(1 - \dfrac{\lambda_{p}}{\mu_{p}}\right) \left( 1 - \delta\right) \left( b_{i,j} + \left(1 - b_{i,j}\right) \left( 1 - f_{sd}\right)\right).
 \end{align}

Given the fact that both battery and relay queues are of finite size, there exists some states with special properties. We now highlight those states and show the impact of their properties on their transition probabilities as follows:
\begin{itemize}
\item States with an empty relay queue $(Q_{sp})$, i.e., $i = 0,\cdots, M$ and $j = 0$ : We set the selection probabilities of those states to 1, i.e., $b_{i,j} = 1,$ in order to prevent wasting any time slots when the SU has the opportunity to access the channel. This, in turn, will effect the transition probabilities as follows
\begin{align}
 P_{i,j \to i,j-1} = 0,
 \end{align}
 \begin{align}
 P_{i,j \to i-1,j} = \left(1 - \dfrac{\lambda_{p}}{\mu_{p}}\right) \left( 1 - \delta\right),
 \end{align}
 \begin{align}
 P_{i,j \to i-1,j-1} = 0.
 \end{align}
\item States with a full relay queue $(Q_{sp})$, i.e., $i = 0,\cdots, M$ and $j = N$ : We set the admission probabilities of those states to 0, i.e., $a_{i,j} = 0,$ since the relay does not have capability to admit any new packets whenever it is full. Hence, transition probabilities at these states are given as
\begin{align}
 P_{i,j \to i+1,j+1} = 0,
 \end{align}
 \begin{align}
 P_{i,j \to i+1,j} = \dfrac{\lambda_{p}}{\mu_{p}} \delta,
 \end{align}
 \begin{align}
 P_{i,j \to i,j+1} = 0.
 \end{align}
\item States with full battery queue at the time when the PU is active, i.e., $i = M$ and $j = 0,\cdots, N - 1$ : We set the energy arrival rate of those states at the time of the PU's activity to 0, i.e., $\delta = 0$, since the energy can not be stored in the battery. Hence, the transition probability form state ($i,j$) to state ($i,j+1$) becomes
\begin{align}
 P_{i,j \to i,j+1} = \dfrac{\lambda_{p}}{\mu_{p}} \left( 1 - f_{pd}\right) f_{ps} a_{i,j}.
 \end{align}
\end{itemize}

 The existence of diagonal transitions, i.e., $P_{i,j \to i-1,j-1}$ and $P_{i,j \to i+1,j+1}$, complicates the solution of the balance equations along with the normalization equation $\sum_{i=0}^{i=M}\sum_{j=0}^{j=N}{\pi_{i,j}} = 1$. Consequently, the product-form solution of the MC is not possible and there is no closed-form expressions for the steady state probabilities. It is worth mentioning that various approximation techniques for the solution of multidimensional MCs have been extensively studied in the literature \cite{9}, e.g., equilibrium point analysis (EPA). However, these approaches rely on approximating the stationary probability distribution of the MC by a unit impulse located at a point in the state space where the system is at equilibrium. Thus, we can not employ these methods to obtain closed-form expressions for the steady state probabilities so that we can optimize the admission and selection probabilities to achieve the maximum SU's service rate subject to stability conditions. Hence, in the next subsection, we propose an algorithm to numerically characterize the stable throughput region.
\subsection{Proposed algorithm for characterizing the stable throughput region numerically}
Tuning the admission and selection probabilities so as to maximize the SU's throughput calls for having closed-form expressions for the steady state distribution of the two-dimensional MC model. Our proposed algorithm is based on discretizing each of the admission and selection probabilities to a number of values, and then investigating the optimal combination of different probabilities that leads to the maximum achievable SU's throughput. Note that the computational complexity of the algorithm increases with the number of states, i.e., as the lengths of $Q_{B}$ and $Q_{sp}$ increase. We will specifically highlight this complexity after presenting our proposed algorithm.

 Algorithm \ref{euclid} presents our proposed method to obtain the maximum achievable throughput of the SU $(\mu_{s}^{*})$ for a given arrival rate of PU $(\lambda_{p})$. For each combination of different admission and selection probabilities, we calculate the SU's achievable throughput as follows. Note that the transition probabilities (\ref{eq5})-(\ref{eq10}) are functions of $\mu_{p}$, where $\mu_{p}$ is a function  of the steady state distribution $(\pi_{i,j})$ from (\ref{eq2}). Therefore, we start by searching for the value of $\mu_{p}$ and $(\pi_{i,j})$ for which both the transition probabilities and (\ref{eq2}) are satisfied, in an iterative manner. More specifically, the steady state distribution is evaluated for each feasible value of $\mu_{p}$, and then the evaluated steady state distribution is used to compute the estimated value of $\mu_{p}$ in Algorithm \ref{euclid} $\left(\bar{\mu_{p}} (Comb_{counter}, \mu_{p})\right)$ from (\ref{eq2}). Next, we choose the value of $\mu_{p}$ which minimizes $\mid \bar{\mu_{p}} (Comb_{counter}, \mu_{p}) - \mu_{p} \mid$. Using the evaluated value of $\mu_{p}$ and the steady state distribution, we compute the SU's achievable throughput for each combination of admission and selection probabilities. Finally, the optimal SU's throughput will be the maximum achievable one among all achievable throughputs for different all combinations.
 
\begin{algorithm}[t!]
\caption{Evaluating the maximum achievable throughput of the SU for a given $\lambda_{p}$ $(\mu_s^* (\lambda_p))$.}\label{euclid}
\begin{algorithmic}
 \STATE Input = $(\lambda_{p}, f_{pd}, f_{ps}, f_{sd}, \delta)$, Output = $\mu_{s}^* (\lambda_{p})$.
 \STATE 1. \textbf{for} $Comb_{counter}$ = 1 : $Comb_{num}$
 \STATE \hspace{1cm} 1) \textbf{for} $\mu_{p}$ = $\lambda_{p} + \epsilon$ : $\theta$ : $f_{pd} + \left(1-f_{pd}\right) f_{ps}$
 \STATE \hspace{2cm} (1)  \textbf{Compute} $\pi_{i,j} (Comb_{counter}, \mu_{p})$.
 \STATE \hspace{2cm} (2)  \textbf{Compute} $\bar{\mu_{p}} (Comb_{counter}, \mu_{p})$ from (\ref{eq2}).
 \STATE \hspace{1cm} 2) \textbf{end for}
 \STATE \hspace{1cm} 3) \textbf{Set} $\mu_{p}^* (Comb_{counter}) = \text{arg}\; \min_{\substack{\mu_{p}}} \; \mid \bar{\mu_{p}} (Comb_{counter}, \mu_{p}) - \mu_{p} \mid$. 
 \STATE \hspace{1cm} 4)  \textbf{Compute} $\mu_{s}(Comb_{counter})$ from (\ref{eq4}).
 \STATE 2. \textbf{end for}
 \STATE 3.  \textbf{Set} $\mu_{s}^{*} (\lambda_p) = \text{arg}\; \max_{\substack{\mu_{s}(Comb_{counter})}} \;\mu_{s}(Comb_{counter})$.
\end{algorithmic}
\end{algorithm}
 The computational complexity of Algorithm \ref{euclid} is determined by two main factors: 1) The number of combinations $(Comb_{num})$ of different selection and admission probabilities, and 2) The increment step size $\theta$ for $\mu_{p}$. Recall that each state $\pi_{i,j}$ is associated with an admission probability $a_{i,j}$ and a selection probability $b_{i,j}$, and there are some states which have deterministic values for either admission or selection probabilities. The number of combinations $Comb_{num}$ can be expressed as $D_{num}^{(M + 1) (N - 1)}$, where $D_{num}$ is the number of discrete values that each of the admission and selection probabilities could take. Therefore, the total number of function evaluations for Algorithm \ref{euclid} is given by $Comb_{num} (2 + c) + 1$, where $c$ denotes the number of iterations of the loop in step 1) and is given by $\dfrac{f_{pd} + \left(1-f_{pd}\right) f_{ps} - \left(\lambda_{p} + \epsilon\right)}{\theta} + 1$. In addition, the total time complexity of Algorithm \ref{euclid} is $\mathcal{O}\left(D_{num}^{(M + 1) (N - 1)}\right)$.

 It is clear that the computational complexity increases exponentially as the queues' lengths ($M$ and $N$) increase. This, in turn, makes the computational time practically infeasible as the lengths of $Q_{sp}$ and $Q_{B}$ become large, and it becomes not possible to even characterize the stable throughput region numerically. Motivated by this sheer computational complexity, we relax the assumption of having state-dependent admission and selection probabilities when $M$ and $N$ are relatively large, and consider that all states, except for those which have either deterministic admission or selection probabilities, have equal admission and selection probabilities. This assumption greatly reduces the computational complexity and also enables us to characterize the stable throughput region for systems with large queues' lengths, as will be shown in the numerical results (Section \ref{sec:num}).

\begin{figure*}[t!]
\centerline{
\subfloat[]{\includegraphics[width=0.8\columnwidth]{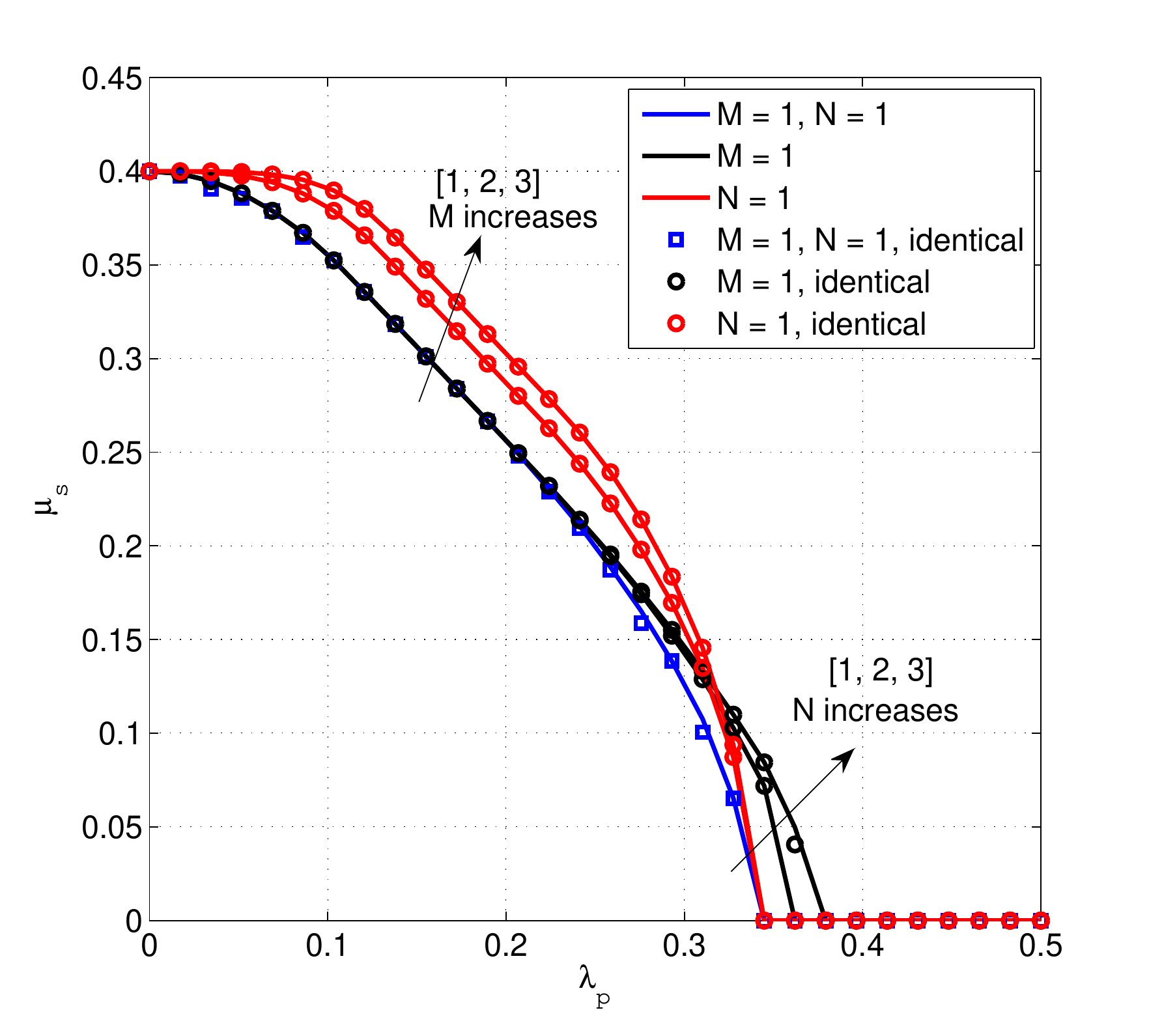}%
\label{fig:3}} \hfil
\subfloat[]{\includegraphics[width=0.8\columnwidth]{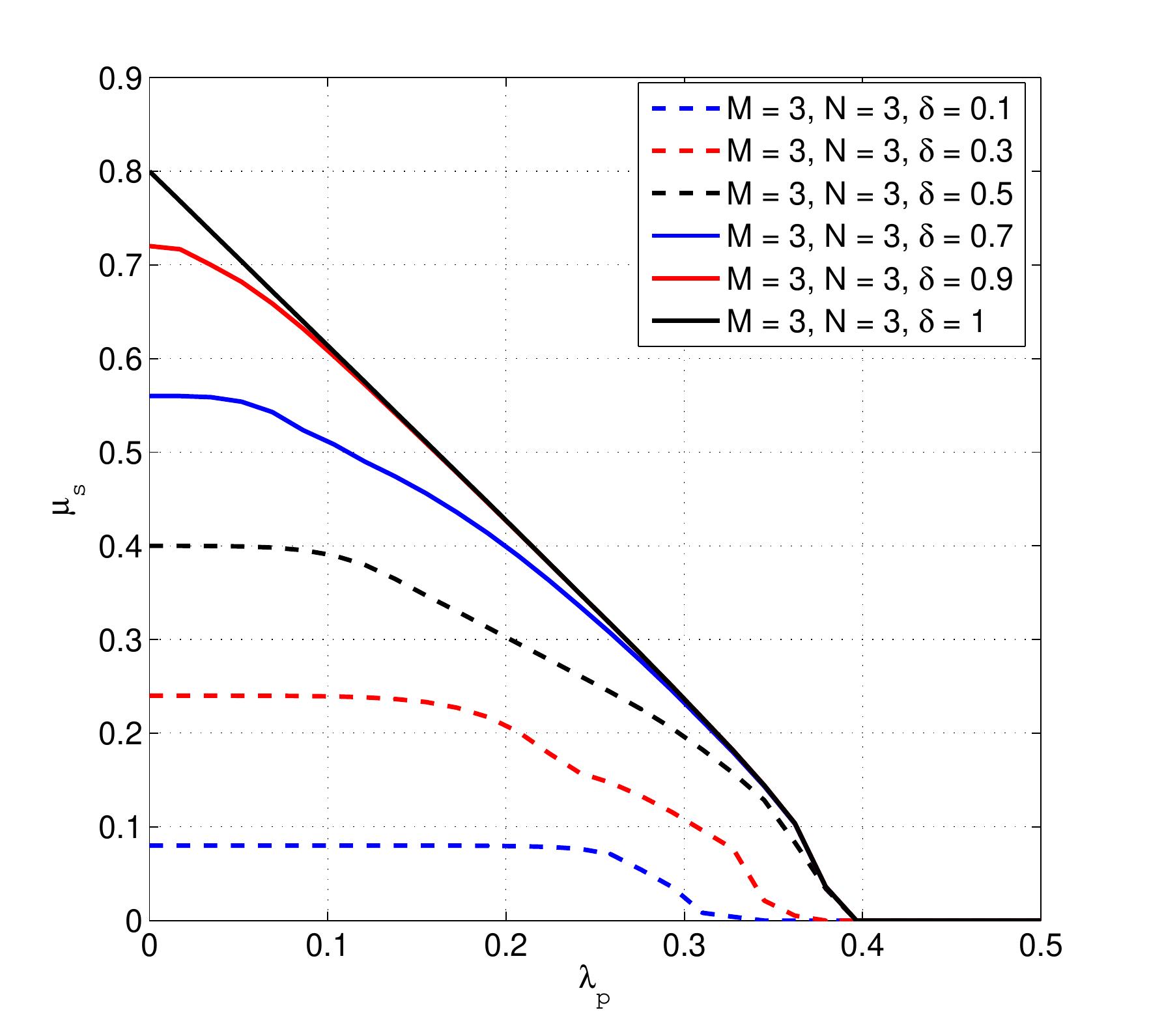}%
\label{fig:4}} \hfil} \caption{Generalized CCRNs with both finite battery and relay queues: (a) the stable throughput region for both systems with different and identical admission and selection probabilities, and (b) the stable throughput region for different values of $\delta$. }
\end{figure*}

 In Fig. \ref{fig:3}, we plot the maximum achievable throughput of the SU $(\mu_{s})$ with respect to the arrival rate of PU's data packets $(\lambda_{p})$, for different combinations of queues' lengths. If it is not stated otherwise, we use the following parameters $f_{pd} = 0.3$, $f_{ps} = 0.4$, $f_{sd} = 0.8$ and $\delta = 0.5$ in the numerical experiments throughout the paper. Here, our main objective is to test the quality of the proposed heuristic scheme of having identical decision variables (i.e., admission and selection probabilities) and relatively large queue lengths. Towards this objective, we quantify the performance loss due to applying the heuristic scheme by comparing its performance to that of Algorithm \ref{euclid}. Interestingly, it is observed that the heuristic scheme achieves the same stable throughput region as the optimal one obtained by Algorithm \ref{euclid}. This, in turn, demonstrates that the optimal stable throughput region for small queues lengths can be achieved using the relatively simple state-independent admission and selection probabilities, except for the boundary states with probabilities either 0 or 1 as explained before. Consequently, the proposed heuristic qualifies as a strong candidate to achieve a near-optimal stable throughput region for large queue lengths (Section \ref{sec:num}). It is also observed that for a fixed relay queue length $N = 1$, increasing the length of the battery queue $(M)$ leads to an expansion in the achievable throughput region. This is explained by the fact that increasing $M$ leads to the ability of storing more energy packets. Therefore, the steady state probability that the battery queue is capable of supporting the transmission of the SU's packets increases and, hence, the maximum SU's achievable throughput increases as well. On the other hand, for a fixed $M = 1$, we observe that as $N$ increases, the maximum sustainable arrival rate of PU's data packets, which is corresponding to a non-zero $\mu_{s}$, increases. This is attributed to the fact that as $N$ increases, the cooperation between the PU and SU becomes more valuable for the PU such that more PU's data packets could be stored at $Q_{sp}$ and served by the SU.

Fig. \ref{fig:4} shows the effect of the arrival rate of the harvested energy packets at the SU on the achievable stable throughput region. Towards that, we fix $N = 3$ and $M = 3$, and plot the stable throughput region for different values of $\delta$ $(\delta =$ $0.1$, $0.3$, $0.5$, $0.7$, $0.9$, and $1$). Increasing $\delta$ leads to having a higher probability of harvesting an energy packet each time slot, which in turn increases the probability of having non-empty $Q_{B}$ at the time when the SU has the ability to access the channel. Therefore, for $\delta \leq 0.5$, it is observed that as $\delta$ increases, both the SU's maximum achievable throughput and PU's maximum sustainable arrival rate increase. On the other hand, for $\delta > 0.5$, increasing $\delta$ leads to a higher maximum achievable throughput for the SU since the maximum sustainable arrival rate of PU's data packets is restricted by the limited size of the relay length.

 In order to investigate the optimization of design parameters at the SU and provide efficient lower computational complexity algorithms for stable throughput region's characterization, we relax the constraint of having two finite queues in the subsequent two sections. Specifically, we focus on two simpler systems, namely, finite battery queue with infinite relay queue in Section \ref{sec:second} and finite relay queue with infinite battery queue in Section \ref{sec:third}.

%

\section{Finite battery queue with infinite relay queue (dominant system $1$)}\label{sec:second}
\subsection{Stability conditions}
Under this setting, we assume that $Q_{B}$ has a maximum finite length $M$, but $Q_{sp}$ may have an infinite queue length. Note that the admission and selection probabilities ($a_{i}$ and $b_{i}$ for $i = 0,\cdots,M$) become only dependent on the state of $Q_{B}$. It is worth nothing that the stability conditions for $Q_{p}$ and $Q_{s}$, given by (\ref{eq3}) and (\ref{eq4}), will reduce to the following expressions

\begin{align}\label{eq11}
\lambda_{p} < f_{pd} + \left(1-f_{pd}\right) f_{ps} \sum_{i=0}^{M}{a_{i} \pi_{i}^{B}},
\end{align}

\begin{align}\label{eq12}
\lambda_{s} < \left(1 - \dfrac{\lambda_{p}}{\mu_{p}}\right) f_{sd} \left(\sum_{i=1}^{M}{b_{i} \pi_{i}^{B}} + \delta b_{0} \pi_{0}^{B}\right),
\end{align} 
respectively, where $\pi_{i}^{B}$ is the steady state probability that $Q_{B}$ has $i$ energy packets at a given time slot. By applying Loynes theorem, the stability condition for $Q_{sp}$ can be derived as follows. A packet is buffered at $Q_{sp}$ if $Q_{p}$ is not empty which happens with probability $1- \dfrac{\lambda_{p}}{\mu_{p}}$. In addition, the packet is not successfully decoded by the destination which happens with probability $1 - f_{pd}$, whereas it is successfully decoded by the SU which happens with probability $f_{ps}$ and is admitted to $Q_{sp}$ which has a probability of $1 - a_{i}$, $i=0,1,\cdots, M$. Thus, $\lambda_{sp}$ is given by
\begin{align}\label{eq13}
\lambda_{sp} = \dfrac{\lambda_{p}}{\mu_{p}} \left(1 - f_{pd}\right) f_{ps} \sum_{i=0}^{M}{a_{i} \pi_{i}^{B}}.
\end{align} 

On the other hand, a packet leaves $Q_{sp}$ if $Q_{p}$ is empty which happens with probability $\dfrac{\lambda_{p}}{\mu_{p}}$, $Q_{B}$ is not empty or it is empty but there is an energy packet arrival, $Q_{sp}$ is selected for transmission which happens with probability $1 - b_{i}$, $i=0,1,\cdots, M$, and the packet is successfully decoded at the destination with probability $f_{sd}$. Therefore, $\mu_{sp}$ and the stability condition for $Q_{sp}$ are given respectively by
\begin{align}\label{eq14}
\mu_{sp} = \left(1- \dfrac{\lambda_{p}}{\mu_{p}}\right) f_{sd} \left(\sum_{i=1}^{M}{\left( 1 - b_{i}\right) \pi_{i}^{B}} + \delta (1 - b_{0}) \pi_{0}^{B}\right),
\end{align}

\begin{align}\label{eq15}
\lambda_{sp} < \mu_{sp}.
\end{align}

\begin{figure}[t!]
\centering
\includegraphics[width=8cm, height= 3cm]{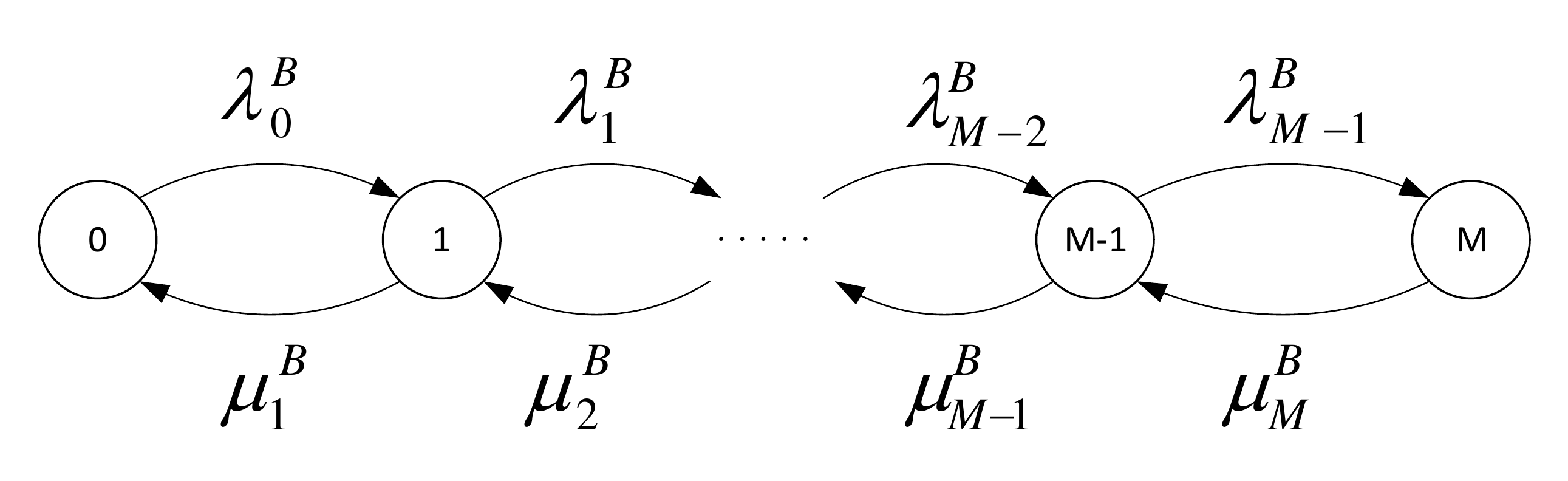}
    \caption{Discrete time MC model for $Q_{B}$ in dominant system $1$.}
     \label{fig:7}
\end{figure}
In order to fully characterize the stability region, we shall now calculate the steady state probabilities of $Q_{B}$ ($\pi_{i}^{B}$ for $ i = 0,\cdots,M$). $Q_{B}$ can be modeled as a discrete time {\rm M}$\mid${\rm M}$\mid$1$\mid${\em M}\footnote{Note that, according to Kendall's notation, the first two {\rm M} letters indicate that the arrival and service processes are Markovian whereas the last {\em M} letter demonstrates that $Q_{B}$ has a maximum finite length $M$.} queue. The MC is shown in Fig. \ref{fig:7} where state $i$ denotes that the number of packets in $Q_B$. Let $\lambda_{i}^{B}$ and $\mu_{i}^{B}$ denote the probability of moving from state $i$ to state $i+1$ and the probability of moving from state $i$ to state $i-1$, respectively. $\lambda_{i}^{B}$ is the probability that $Q_{p}$ is not empty and an energy packet arrives at $Q_{B}$. On the other hand, $\mu_{i}^{B}$ is the probability that $Q_{p}$ is non-empty and there is no arriving energy packet. Thus, using the balance equations, the steady state probabilities of $Q_{B}$ are given by

 \begin{align} \label{eq16}
 \pi_{i+1}^{B} = \dfrac{\delta\dfrac{\lambda_{p}}{\mu_{p}}}{\left(1 - \dfrac{\lambda_{p}}{\mu_{p}}\right) \left(1 - \delta\right)} \pi_{i}^{B},
 \end{align}
where $i=0,1,\cdots,M - 1$. Applying the normalization condition

\begin{align}\label{eq17}
\sum_{i=0}^{M}{\pi_{i}^{B}} = 1,
\end{align}
along with (\ref{eq16}), the steady state distribution of $Q_{B}$ can be completely characterized.

The main results of this subsection can be summarized in the following proposition.
\begin{proposition}\label{pro:1} 
Given $a_{i}$ and $b_{i}$, $i=0,1,\cdots, M$, the dominant system $1$ is stable if the arrival rates of $Q_{p}$, $Q_{s}$ and $Q_{B}$ satisfy the following conditions:
\begin{align*}
\lambda_{p} < f_{pd} + \left(1-f_{pd}\right) f_{ps} \sum_{i=0}^{M}{a_{i} \pi_{i}^{B}},
\end{align*}
\begin{align*}
\lambda_{s} < \left(1 - \dfrac{\lambda_{p}}{\mu_{p}}\right) f_{sd} \left(\sum_{i=1}^{M}{b_{i} \pi_{i}^{B}} + \delta b_{0} \pi_{0}^{B}\right),
\end{align*}
\begin{align*}
\lambda_{sp} < \mu_{sp},
\end{align*}
where $\lambda_{sp}$ and $\mu_{sp}$ are given by (\ref{eq13}) and (\ref{eq14}), respectively. $\pi_{i}^{B}$, $i=0,1,\cdots, M$, can be obtained by solving equations (\ref{eq16}) and (\ref{eq17}).
\end{proposition} 

\subsection{SU's throughput maximization problem}
The SU's service rate maximization problem for dominant system $1$ can be formulated as
 \begin{equation}\label{eq23}
 \begin{aligned}
& \textbf{P1}: & & \max_{\substack{a_{i},b_{i},\pi_{i}^{B},\mu_{p}}} \; \; \left(1 - \dfrac{\lambda_{p}}{\mu_{p}}\right) f_{sd} \left(\sum_{i=1}^{M}{b_{i} \pi_{i}^{B}} + \delta b_{0} \pi_{0}^{B}\right) \\
&\text{s.t.} &&\lambda_{p} < \mu_{p}, \\
&&& \mu_{p} = f_{pd} + \left(1-f_{pd}\right) f_{ps} \sum_{i=0}^{M}{a_{i} \pi_{i}^{B}},  \\
&&& \lambda_{sp} < \mu_{sp}, \\
&&& 0 \leq \pi_{i}^{B}, a_{i}, b_{i} \leq 1,\;i=0, \cdots,M, \\
&&& (\ref{eq16}),(\ref{eq17}),
\end{aligned}
\end{equation}
where $\lambda_{sp}$ and  $\mu_{sp}$ are given by (\ref{eq13}) and (\ref{eq14}), respectively.

It is worth nothing that \textbf{P1} is a non-convex optimization problem. However, we exploit the problem structure in order to transform it into a linear program. More specifically, by defining the new variables 

\begin{align}\label{eq24}
x_{i} = a_{i} \pi_{i}^{B}, y_{i} = b_{i} \pi_{i}^{B},\; i=0, \cdots,M,
\end{align}
\textbf{P1} reduces into a linear program for a given $\mu_{p}$ as follows. First, we have the following constraints on the new defined variables

\begin{align}\label{eq25}
0 \leq x_{i}, y_{i} \leq \pi_{i}^{B},\; i=0, \cdots,M.
\end{align}

Second, we can rewrite the constraint in (\ref{eq15}) as
\begin{align}\label{eq26}
\sum_{i=0}^{M}{x_{i}} < \dfrac{f_{sd}\left(\mu_{p} - \lambda_{p}\right)}{\lambda_{p}f_{ps}\left(1 - f_{pd}\right)} \left(\sum_{i=1}^{M}{\left(\pi_{i}^{B} - y_{i}\right)} + \delta\left( \pi_{0}^{B} - y_{0}\right)\right).
\end{align}

Finally, by substituting with the new defined variables into the objective function and the remaining constraints, \textbf{P1} turns out to be a linear program for a given $\mu_{p}$ and can be expressed as follows

\begin{equation}\label{eq27}
 \begin{aligned}
& \textbf{P1}^{\ast}: & & \max_{\substack{x_{i},y_{i},\pi_{i}^{B}}} \; \; \left(1 - \dfrac{\lambda_{p}}{\mu_{p}}\right) f_{sd} \left(\sum_{i=1}^{M}{y_{i}} + \delta y_{0}\right) \\
&\text{s.t.} &&\mu_{p} = f_{pd} + \left(1-f_{pd}\right) f_{ps} \sum_{i=0}^{M}{x_{i}}, \\
&&& 0 \leq \pi_{i}^{B} \leq 1,\;i=0, \cdots,M, \\ 
&&& (\ref{eq16}),(\ref{eq17}),(\ref{eq25}), (\ref{eq26}).
\end{aligned}
\end{equation}

From (\ref{eq11}), the feasible values of $\mu_{p}$ over which the linear program is solved are given by
\begin{align}\label{Mp range for P2}
\text{max} (\lambda_{p}, f_{pd}) \leq \mu_{p} \leq f_{pd} + \left(1-f_{pd}\right) f_{ps}.
\end{align}

\begin{algorithm}[t!]
\caption{Evaluating the maximum achievable throughput of the SU for a given $\lambda_{p}$ (dominant system $1$).}\label{Algorithm2}
\begin{algorithmic}
 \STATE Input = $(\lambda_{p}, f_{pd}, f_{ps}, f_{sd}, \delta)$, Output = $\mu_{s}^* (\lambda_{p})$.
 \STATE 1. \textbf{for} $\mu_{p}$ = $\text{max} (\lambda_{p}, f_{pd})$: $\theta$ : $f_{pd} + \left(1-f_{pd}\right) f_{ps}$
 \STATE \hspace{1cm} 1) Compute $\mu_{s}^{*}(\mu_{p})$ from $\textbf{P1}^{\ast}$.
 \STATE 2. \textbf{end for}
 \STATE 3. \textbf{Set} $\mu_{p}^{*} = \text{arg}\; \max_{\substack{\mu_{p}}} \;\mu_{s}^{*}(\mu_{p})$.
 \STATE 4. \textbf{Set} $\mu_{s}^{*} (\lambda_p) = \mu_{s}^{*}(\mu_{p}{*})$.
\end{algorithmic}
\end{algorithm}
\begin{figure*}[t!]
\centerline{
\subfloat[]{\includegraphics[width=0.8\columnwidth]{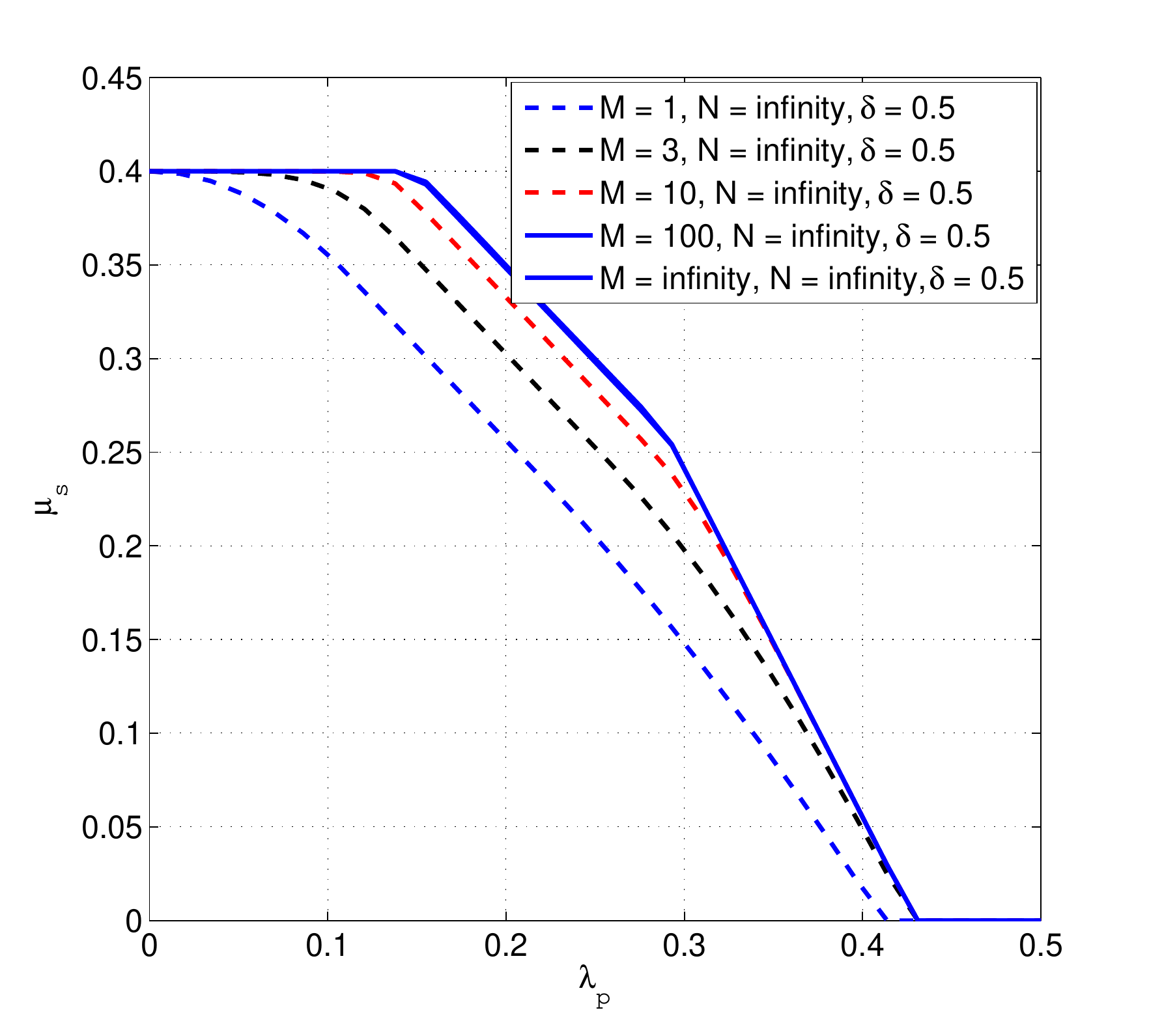}%
\label{fig:8}} \hfil
\subfloat[]{\includegraphics[width=0.8\columnwidth]{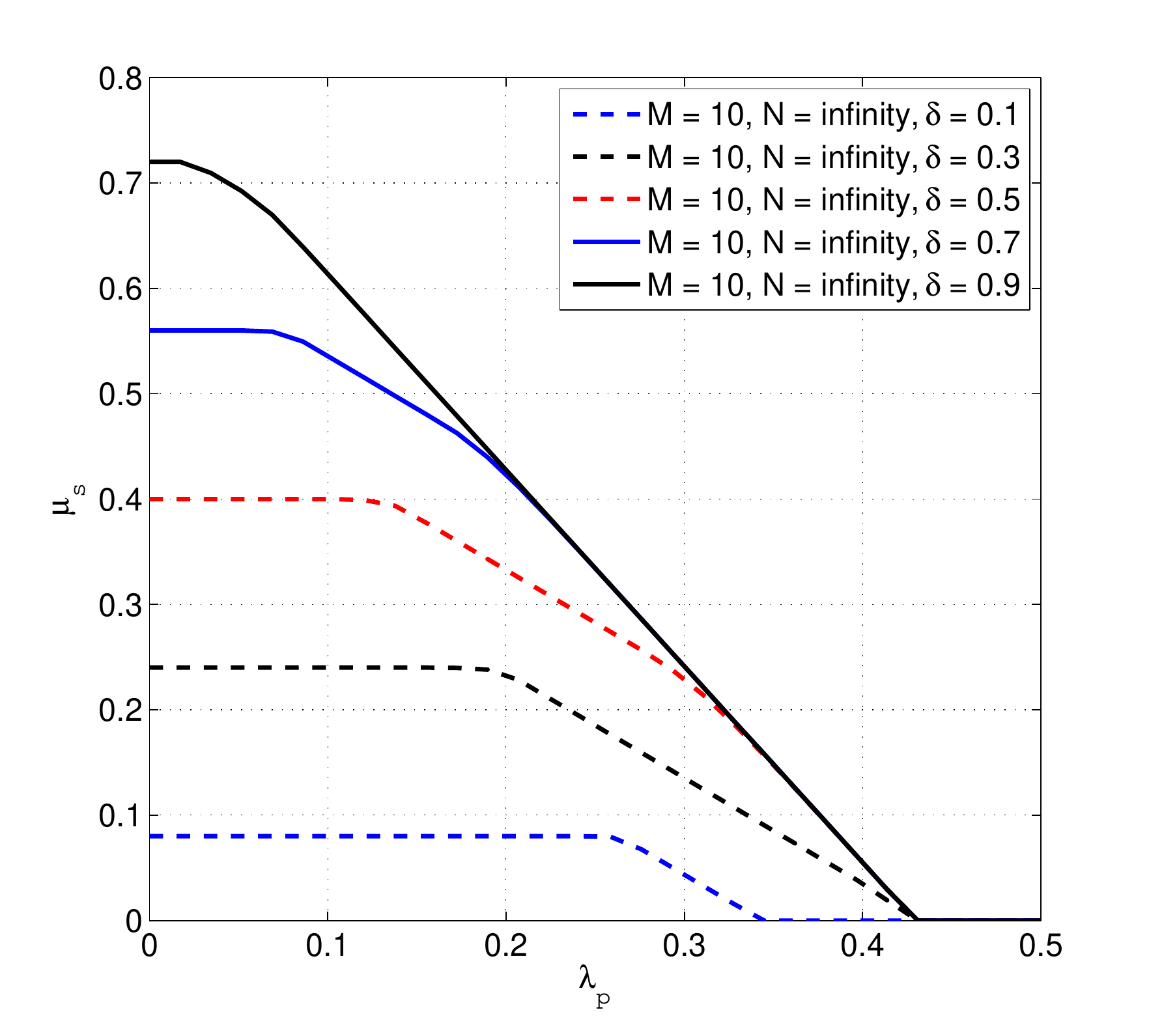}%
\label{fig:9}} \hfil} \caption{Dominant system $1$: (a) the stable throughput region for different values of $M$, and (b) the stable throughput region for different values of $\delta$.}
\end{figure*}

We now show how to obtain the maximum achievable throughput of the SU for a given $\lambda_{p}$ using Algorithm \ref{Algorithm2}. For a given $\lambda_{p}$, the feasible range of $\mu_{p}$'s values is defined by (\ref{Mp range for P2}). For each feasible $\mu_{p}$, the maximum SU's achievable throughput is obtained by solving $\textbf{P1}^{\ast}$. Afterwards, we search for the optimal value of $\mu_{p}$ which achieves the maximum throughput in the feasible range of $\mu_{p}$'s values. Finally, the obtained optimal value of $\mu_{p} (\mu_{p}^{*})$ is used to evaluate the optimal SU's achievable throughput $(\mu_{s}^{*} (\lambda_p))$. We use standard optimization tools, e.g., CVX \cite{10}, to obtain the optimal solution.

  Fig. \ref{fig:8} compares the achievable stable throughput region of a baseline cooperative cognitive radio with infinite size battery and relay queues with that of $\textbf{P1}^*$ for different values of maximum battery queue length $(M =$ $1$, $3$, $10$ and $100$). As expected, the stable throughput region obtained by $\textbf{P1}^*$ expands as $M$ increases. Increasing $M$ decreases the probability of $Q_{B}$ being empty and the drop of energy packets due to overflow. Therefore, the service rate of the SU increases with $M$. It is observed also that the system with finite battery length $M = 100$ achieves the same stable throughput region as in the scenario of having infinite battery length. Hence, in practical systems, a finite battery capacity of a sufficient size $(M = 100)$ is enough to reap the same level of benefits of a system with infinite size battery.
 
  In Fig. \ref{fig:9}, we demonstrate the effect of the arrival rate of the harvested energy packets at the SU on the stable throughput region for dominant system $1$ $(\textbf{P1}^*)$. To this end, we fix $M = 10$ and plot the stable throughput region for different values of $\delta$ $(\delta =$ $0.1$, $0.3$, $0.5$, $0.7$ and $0.9$). It is observed that as the average arrival rate of the harvested energy packets increases, the throughput region expands. This happens since as the average arrival rate of harvested energy packets increases the likelihood that $Q_{B}$ is empty decreases. That, in turn, manages the SU to achieve larger service rate $(\mu_{s})$ for a given PU packet arrival rate $(\lambda_{p})$.


\section{Finite relay queue with infinite battery queue (dominant system $2$)}\label{sec:third}
\subsection{Stability conditions}
Under this setting, we assume that $Q_{sp}$ remains with finite length $N$, but $Q_{B}$ becomes an infinite size queue. Although the effects of finite size relay queue were studied in \cite{5}, it was implicitly assumed that the system has no energy limitation, i.e., the SU always has energy packets to transmit whenever it has the opportunity to access the channel. On the contrary, in this subsection, we focus on the more interesting practical scenario of having a limited-energy system. The energy limitation is characterized through the fact that there is a non-zero probability of having an empty $Q_{B}$ for a certain range of $\lambda_{p}$'s values. Thus, the number of energy packets inside $Q_{B}$ will never grow to infinity for such range of $\lambda_{p}$'s values and we will have a limited-energy system.

In this scenario, the admission and selection probabilities ($a_{j}$ and $b_{j}$ for $j = 0,\cdots,N$) become only dependent on the state of $(Q_{sp})$. Recall that $a_{N} = 0$ to prevent $Q_{sp}$ from admitting any overheard PU's packet when it is full, and $b_{0} = 1$ to prevent allocating any transmission time slots for $Q_{sp}$ when it is empty. The stability condition for $Q_{p}$, given by (\ref{eq3}), will reduce to the following expression 
\begin{align}\label{eq18}
\lambda_{p} < f_{pd} + \left(1-f_{pd}\right) f_{ps} \sum_{j=0}^{N}{a_{j} \pi_{j}^{sp}},
\end{align}
where $\pi_{j}^{sp}$ is the steady state probability that $Q_{sp}$ has $j$ packets at a given time slot. The probability of moving from state $i$ to state $i+1$ at $Q_{B}$ is the probability that $Q_p$ is non-empty and there is an energy packet arrival. Thus, the probability of moving from state $i$ to state $i+1$ at $Q_{B}$ can be expressed as 
\begin{align}
\lambda_{B} = \delta \dfrac{\lambda_{p}}{\mu_{p}}.
\end{align}

On the other hand, due to applying the dominant system approach, the probability of moving from state $i$ to state $i-1$ at $Q_{B}$ is the probability that $Q_p$ is empty and there is no energy packet arrival. Therefore, the probability of moving from state $i$ to state $i-1$ at $Q_{B}$ can be expressed as 
\begin{align}
\mu_{B} = \left(1 - \dfrac{\lambda_{p}}{\mu_{p}}\right) \left(1 - \delta\right).
\end{align}

Depending on the relationship between $\lambda_{B}$ and $\mu_{B}$, the energy available at $Q_{B}$ is determined. If $\lambda_{B}$ is strictly less than $\mu_{B}$, the probability that $Q_{B}$ is capable of supporting the transmission of the SU's packet will be $\delta /\left(1 - \dfrac{\lambda_{p}}{\mu_{p}}\right)$, and it is the sum of the two probabilities: i) the probability of having non-empty $Q_{B}$ which is $\lambda_{B} / \mu_{B}$ and ii) the probability of having an empty $Q_{B}$ but there is an energy packet arrival which is $\delta \left(1 - \dfrac{\lambda_{B}}{\mu_{B}}\right)$. On the other hand, when $\mu_{B}$ is less than or equal to $\lambda_{B}$, the number of energy packets inside $Q_{B}$ will grow to infinity and the system will be considered as being equipped with unlimited-energy supply. Thus, the stability condition for $Q_{s}$, given by (\ref{eq4}), reduces to the following expression
\begin{align}\label{eq19}
\lambda_{s} < 
 \begin{cases}
 \left(1 - \dfrac{\lambda_{p}}{\mu_{p}}\right) f_{sd} \dfrac{\delta}{1 - \dfrac{\lambda_{p}}{\mu_{p}}} \sum_{j=0}^{N}{b_{j} \pi_{j}^{sp}},\; \; \lambda_{B} < \mu_{B} \\
\left(1 - \dfrac{\lambda_{p}}{\mu_{p}}\right) f_{sd} \sum_{j=0}^{N}{b_{j} \pi_{j}^{sp}},\; \; \lambda_{B} \geq \mu_{B}.
 \end{cases}
\end{align}
%
\begin{figure}[t!]
\centering
\includegraphics[width=8 cm, height= 3cm]{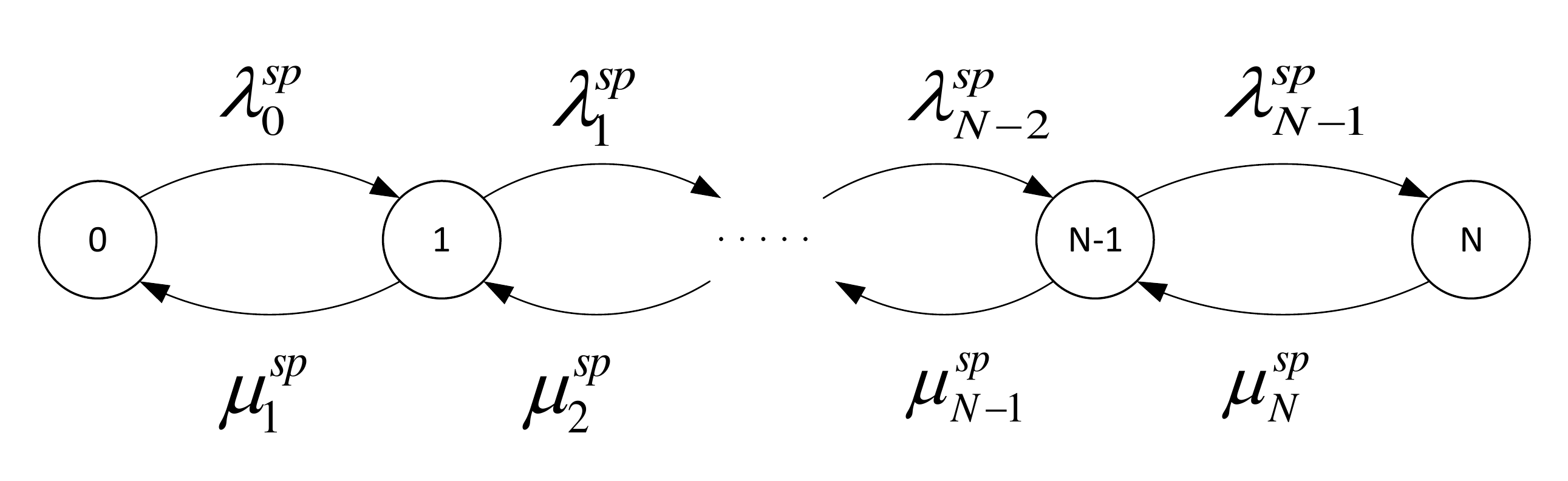}
    \caption{Discrete time MC model for $Q_{sp}$ in dominant system $2$.}
     \label{fig:10}
\end{figure}
Next, we calculate the steady state distribution of $Q_{sp}$. $Q_{sp}$ can be modeled as a discrete time {\rm M}$\mid${\rm M}$\mid$1$\mid$ $N$ queue. The MC is shown in Fig. \ref{fig:10} where state $j$ denotes that the number of packets in $Q_{sp}$ is $j$. Let $\lambda_{j}^{sp}$ and $\mu_{j}^{sp}$ denote the probability of moving from state $j$ to state $j+1$ and the probability of moving from state $j$ to state $j-1$, respectively. $\lambda_{j}^{sp}$ is the probability that $Q_{p}$ is not empty, the packet is not successfully decoded by the destination, whereas it is successfully decoded by the SU and is admitted to $Q_{sp}$. On the other hand, $\mu_{j}^{sp}$ is the probability that $Q_{p}$ is empty, $Q_{B}$ is capable of supporting the transmission of the SU's packet, $Q_{sp}$ is selected for transmission and the packet is successfully decoded at the destination. Thus, using the balance equations, the steady state probabilities of $Q_{sp}$ are given by
 \begin{align} \label{eq21}
 \pi_{j+1}^{sp} =\dfrac{\lambda_{j}^{sp}}{\mu_{j+1}^{sp}} \pi_{j}^{sp},
 \end{align}
where $j=0,1,\cdots,N - 1$, and $\lambda_{j}^{sp}$ and $\mu_{j+1}^{sp}$ are given respectively by
\begin{align} 
 \lambda_{j}^{sp} =\dfrac{\lambda_{p}}{\mu_{p}}f_{ps}\left(1 - f_{pd}\right)a_{j},
 \end{align}
 \begin{equation} 
\mu_{j+1}^{sp} = 
  \begin{cases}
 \delta f_{sd} \left(1 - b_{j+1}\right),\; \; \lambda_{B} < \mu_{B} \\
 \left(1 - \dfrac{\lambda_{p}}{\mu_{p}}\right)f_{sd}\left(1 - b_{j+1}\right),\; \; \lambda_{B} \geq \mu_{B}.
 \end{cases}
\end{equation}

Applying the normalization condition
\begin{align}\label{eq22}
\sum_{i=0}^{N}{\pi_{j}^{sp}} = 1,
\end{align}
along with (\ref{eq21}), the steady state distribution of $Q_{sp}$ can be completely characterized. In the next subsection, we formulate the stable throughput region optimization problem and discuss its solution.

The main results of this subsection can be summarized in the following proposition.
\begin{proposition}\label{pro:2} 
Given $a_{j}$ and $b_{j}$, $i=0,1,\cdots, N$, the dominant system $2$ is stable if the arrival rates of $Q_{p}$ and $Q_{s}$ satisfy the following conditions:
\begin{align*}
\lambda_{p} < f_{pd} + \left(1-f_{pd}\right) f_{ps} \sum_{j=0}^{N}{a_{j} \pi_{j}^{sp}},
\end{align*}
\begin{align*}
\lambda_{s} < 
 \begin{cases}
 \left(1 - \dfrac{\lambda_{p}}{\mu_{p}}\right) f_{sd} \dfrac{\delta}{1 - \dfrac{\lambda_{p}}{\mu_{p}}} \sum_{j=0}^{N}{b_{j} \pi_{j}^{sp}},\; \; \lambda_{B} < \mu_{B} \\
\left(1 - \dfrac{\lambda_{p}}{\mu_{p}}\right) f_{sd} \sum_{j=0}^{N}{b_{j} \pi_{j}^{sp}},\; \; \lambda_{B} \geq \mu_{B}.
 \end{cases},
\end{align*}
where $\pi_{j}^{sp}$, $j=0,1,\cdots, N$, can be obtained by solving equations (\ref{eq21}) and (\ref{eq22}).
\end{proposition}

\subsection{SU's throughput maximization problem}
The SU's service rate maximization problem of dominant system $2$ for a given $\lambda_{p}$, when $\lambda_{B} < \mu_{B}$, can be formulated as
 \begin{equation}\label{eq28}
 \begin{aligned}
& \textbf{P2}: & & \max_{\substack{a_{j},b_{j},\pi_{j}^{sp},\mu_{p}}} \; \; \delta f_{sd} \sum_{j=0}^{N}{b_{j} \pi_{j}^{sp}} \\
&\text{s.t.} &&\lambda_{p} < \mu_{p}, \\
&&& \mu_{p} = f_{pd} + \left(1-f_{pd}\right) f_{ps} \sum_{j=0}^{N}{a_{j} \pi_{j}^{sp}}, \\
&&& \pi_{j+1}^{sp} =\dfrac{\dfrac{\lambda_{p}}{\mu_{p}}f_{ps}\left(1 - f_{pd}\right)a_{j}}{f_{sd} \delta \left(1 - b_{j+1}\right)} \pi_{j}^{sp},\; j=0, \cdots,N-1,\\
&&& \sum_{i=0}^{N}{\pi_{j}^{sp}} = 1,\\
&&& a_{N} = 0,\; b_{0} = 1, \\
&&& 0 \leq a_{j}, b_{j}, \pi_{j}^{sp} \leq 1,\; j=0, \cdots,N.
\end{aligned}
\end{equation}

By inspecting \textbf{P2}, we can easily see that it is a non-convex optimization problem. However, similar to \textbf{P1}, \textbf{P2}'s structure can be exploited to transform it into a linear program for a given $\mu_{p}$. Recall that, from (\ref{eq18}), the feasible values of $\mu_{p}$ over which the linear program runs are given by (\ref{Mp range for P2}).

 By defining the new variables 
\begin{align}\label{eq29}
x_{j} = a_{j} \pi_{j}^{sp}, y_{j} = b_{j} pi_{j}^{sp},\; j=0, \cdots,N,
\end{align}
\textbf{P2} reduces into a linear program for a given $\mu_{p}$ as follows. First, we have the following constraints on the new defined variables

\begin{align}\label{eq30}
0 \leq x_{j}, y_{j} \leq \pi_{j}^{sp},\; j=0, \cdots,N.
\end{align}

Second, we can rewrite the constraint in (\ref{eq21}) as
\begin{align}\label{eq31}
\pi_{j+1}^{sp} - y_{j+1} = \dfrac{\lambda_{p} f_{ps}\left(1 - f_{pd}\right)}{\mu_{p} f_{sd} \delta} x_{j},\; j=0, \cdots,N - 1.
\end{align}

Finally, by substituting the new defined variables into the objective function and the remaining constraints, \textbf{P2} turns out to be a linear program for a given $\mu_{p}$ and can be expressed as follows

\begin{equation}\label{eq32}
 \begin{aligned}
& \textbf{P2}^{\ast}: & & \max_{\substack{x_{j},y_{j},\pi_{j}^{sp}}} \; \; \delta f_{sd} \sum_{j=0}^{N}{y_{j}} \\
&\text{s.t.} &&\mu_{p} = f_{pd} + \left(1-f_{pd}\right) f_{ps} \sum_{j=0}^{N}{x_{j}}, \\
&&& x_{N} = 0,\; y_{0} = \pi_{0}^{sp}, \\
&&& 0 \leq \pi_{j}^{sp} \leq 1,\;j=0, \cdots,N, \\ 
&&& \sum_{i=0}^{N}{\pi_{j}^{sp}} = 1,\\
&&& (\ref{eq30}),(\ref{eq31}).
\end{aligned}
\end{equation}

Following the same approach applied to \textbf{P2}, the SU's service rate maximization problem of dominant system $2$ for a given $\lambda_{p}$, when $\lambda_{B} \geq \mu_{B}$, can be formulated as
\begin{equation}
 \begin{aligned}
& \textbf{P3}: & & \max_{\substack{x_{j},y_{j},\pi_{j}^{sp}}} \; \; \left(1 - \dfrac{\lambda_{p}}{\mu_{p}}\right) f_{sd} \sum_{j=0}^{N}{y_{j}} \\
&\text{s.t.} &&\mu_{p} = f_{pd} + \left(1-f_{pd}\right) f_{ps} \sum_{j=0}^{N}{x_{j}}, \\
&&& \pi_{j+1}^{sp} - y_{j+1} = \dfrac{\lambda_{p} f_{ps}\left(1 - f_{pd}\right)}{ f_{sd} \left(\mu_{p} - \lambda_{p}\right)} x_{j},\; j=0, \cdots,N - 1,\\
&&& \sum_{i=0}^{N}{\pi_{j}^{sp}} = 1,\\
&&& x_{N} = 0,\; y_{0} = \pi_{0}^{sp}, \\
&&& 0 \leq \pi_{j}^{sp} \leq 1,\;j=0, \cdots,N, \\ 
&&& (\ref{eq30}).
\end{aligned}
\end{equation}

 \begin{algorithm}[t!]
\caption{Evaluating the maximum achievable throughput of the SU for a given $\lambda_{p}$ (dominant system $2$).}\label{Algorithm3}
\begin{algorithmic}
 \STATE Input = $(\lambda_{p}, f_{pd}, f_{ps}, f_{sd}, \delta)$, Output = $\mu_{s}^* (\lambda_{p})$.
 \STATE 1. \textbf{for} $\mu_{p}$ = $\text{max} (\lambda_{p}, f_{pd})$: $\theta$ : $f_{pd} + \left(1-f_{pd}\right) f_{ps}$
 \STATE \hspace{1cm} 1) \textbf{if} ($\lambda_{B} < \mu_{B}$)
 \STATE \hspace{2cm}(1) Compute $\mu_{s}^{*}(\mu_{p})$ from $\textbf{P2}^{\ast}$.
 \STATE \hspace{1cm} 2) \textbf{else} 
 \STATE \hspace{2cm}(2) Compute $\mu_{s}^{*}(\mu_{p})$ from $\textbf{P3}$.
 \STATE \hspace{1cm} 3) \textbf{end if} 
 \STATE 2. \textbf{end for}
 \STATE 3. \textbf{Set} $\mu_{p}^{*} = \text{arg}\; \max_{\substack{\mu_{p}}} \;\mu_{s}^{*}(\mu_{p})$.
 \STATE 4. \textbf{Set} $\mu_{s}^{*} (\lambda_p) = \mu_{s}^{*}(\mu_{p}{*})$.
\end{algorithmic}
\end{algorithm}

We now summarize how to obtain the SU's achievable throughput for a given $\lambda_{p}$ using Algorithm \ref{Algorithm3}. For a given $\lambda_{p}$, the feasible range of $\mu_{p}$'s values is defined by (\ref{Mp range for P2}). For each feasible $\mu_{p}$, the maximum SU's achievable throughput is either obtained by solving $\textbf{P2}^{\ast}$ if $\lambda_{B} < \mu_{B}$ or by solving $\textbf{P3}$ if $\lambda_{B} \geq \mu_{B}$. Finally, we search for the optimal value of $\mu_{p}$ which achieves the maximum throughput in the feasible range of $\mu_{p}$'s values. Note that increasing $M$ in Algorithm \ref{Algorithm2} ($N$ in Algorithm \ref{Algorithm3}) leads to a higher number of optimization variables in $\textbf{P1}^{\ast}$ ($\textbf{P2}^{\ast}$ and \textbf{P3}). Owing to the convexity of $\textbf{P1}^{\ast}$, $\textbf{P2}^{\ast}$ and \textbf{P3}), the computational complexity of Algorithm \ref{Algorithm2} or Algorithm \ref{Algorithm3} does not significantly increase as a function of $M$ or $N$, respectively, and thus both Algorithms have a much lower computational complexity compared to that of Algorithm \ref{euclid}. Particularly, the total time complexity of both Algorithm \ref{Algorithm2} and Algorithm \ref{Algorithm3} is $\mathcal{O}(1)$.

\begin{figure*}[t!]
\centerline{
\subfloat[]{\includegraphics[width=0.8\columnwidth]{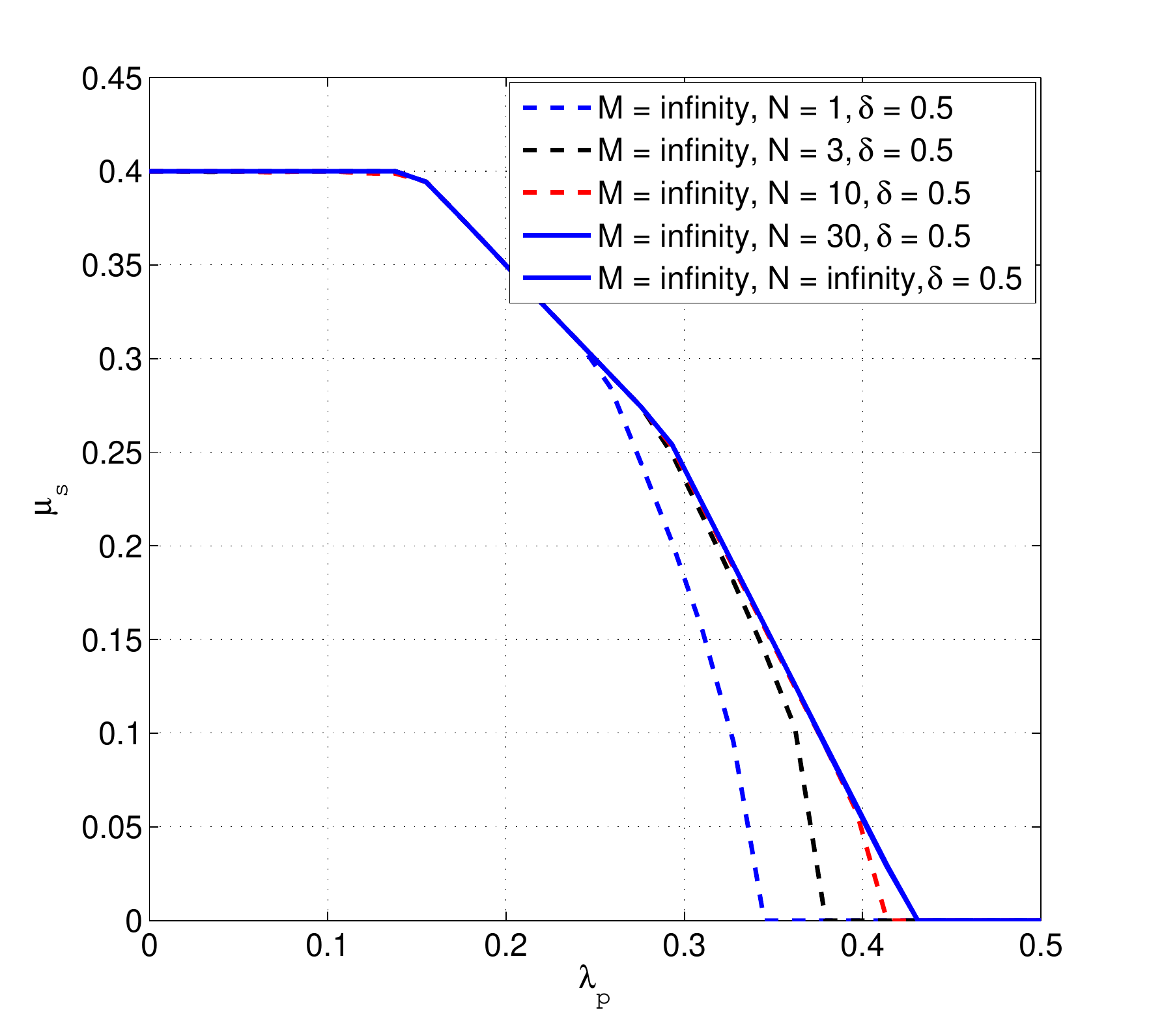}%
\label{fig:11}} \hfil
\subfloat[]{\includegraphics[width=0.8\columnwidth]{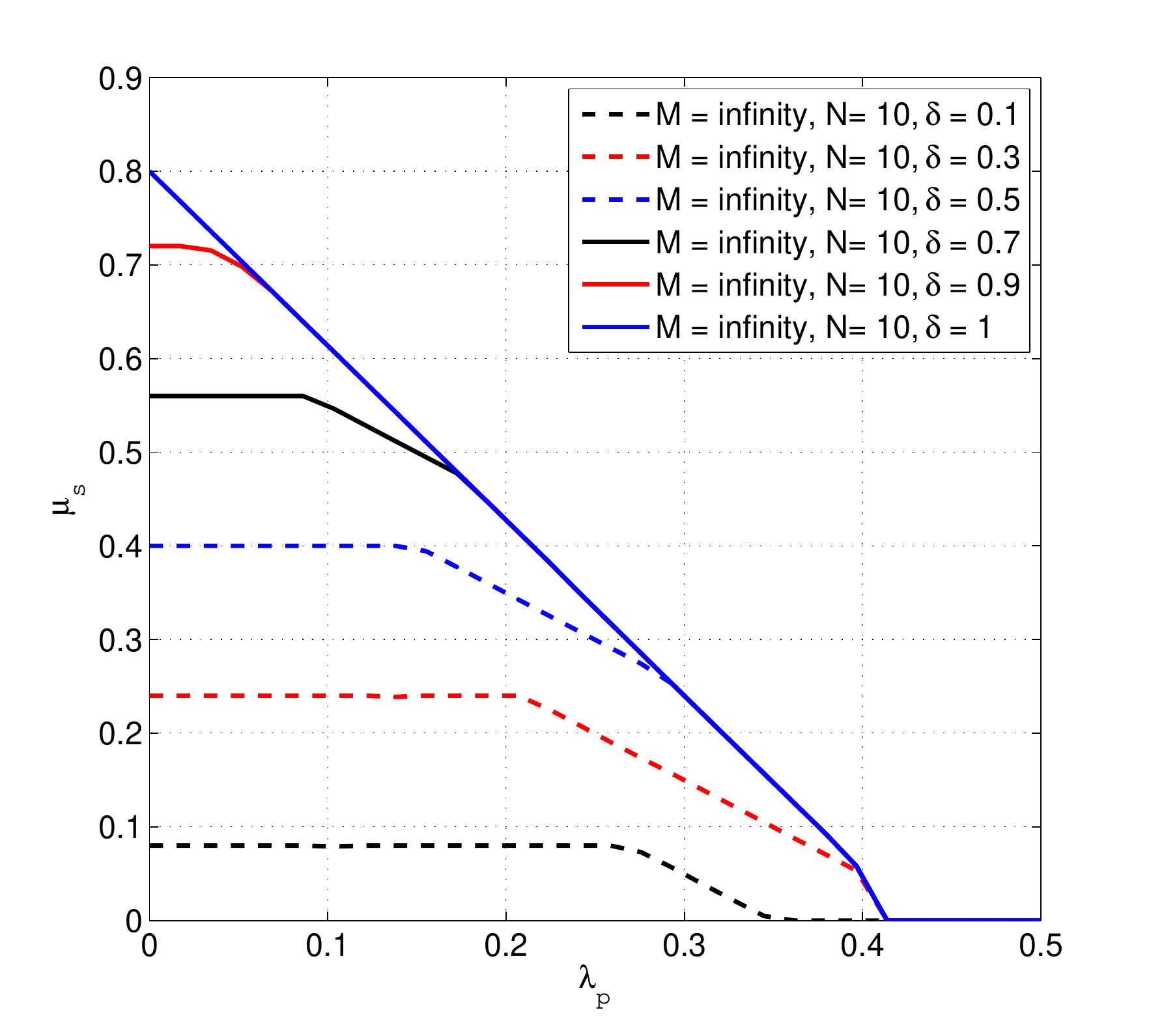}%
\label{fig:12}} \hfil} \caption{Dominant system $2$: (a) the stable throughput region for different values of $N$, and (b) the stable throughput region for different values of $\delta$.}
\end{figure*}

Fig. \ref{fig:11} compares the achievable stable throughput region of dominant system $2$ with that of the baseline cooperative cognitive radio with infinite battery and relay queue sizes, for different values of $Q_{sp}$ length $(N =$ $1$, $3$, $10$ and $30$). On the contrary to the achievable throughput region by dominant system $1$, we observe here two different regimes due to the increase of $Q_{sp}$'s length. The first regime is represented by the range $ \lambda_{p} \leq 0.25$, where all systems with different lengths achieve the same maximum SU's achievable throughput. This happens due to the fact that, for small values of $\lambda_p$, the likelihood that the SU has the opportunity to access the channel is relatively high. Thus, the capability of $Q_{B}$ to support the transmission of the SU's packets will serve as the bottleneck of the maximum achievable SU's throughput. Since the capability of $Q_{B}$ to support the transmission of the SU's packets is defined by $Q_{B}$'s length $(M)$ and $\delta$ ($M$ and $\delta$ are the same for all plotted systems), we observe that the SU's achievable throughput is the same for all systems. On the other hand, the second regime is represented by the range $\lambda_{p} > 0.25$, where the value of $N$ defines the maximum sustainable arrival rate of PU's data packets, which is corresponding to a non-zero $\mu_{s}$. Increasing $N$ reduces the time slots needed by the PU to serve its own data packets and, hence, increases the likelihood that the SU has the chance to access the channel. Therefore, it is observed that the maximum sustainable arrival rate of PU's data packets, which is corresponding to a non-zero $\mu_{s}$, increases with $N$. Finally, we observe that the system with finite relay length $N = 30$ achieves the same stable throughput region as in the scenario of having infinite relay length. Hence, in realistic systems, a finite relay queue of a size $N = 30$ is enough to reap the same level of benefits of a system with infinite size relay queue.

Fig. \ref{fig:12} shows the stable throughput region achieved by dominant system $2$, for different values of the average arrival rate of the harvested energy packets ($\delta$) and $N = 10$. Increasing $\delta$ promotes the probability that $Q_{B}$ is capable of supporting the transmission of the SU's packets. Therefore, it is observed that as $\delta$ increases, the achievable throughput region expands.

Finally we investigate the range of $\lambda_{p}$'s values, over which there exists a non-zero probability of having an empty $Q_{B}$ and $\textbf{P2}^{\ast}$ is used to characterize the maximum achievable throughput. Towards that, we introduce the energy-limited cooperative cognitive radio networks with finite relay queue and infinite battery queue, in which we impose an energy limitation constraint to ensure that there is always a non-zero probability of having an empty $Q_{B}$. More specifically, the energy limitation constraint guarantees that the energy arrival rate at $Q_{B}$ is strictly less than its service rate. Thus, the number of energy packets inside $Q_{B}$ will never grow to infinity. The stable throughput region of such networks is characterized by solving $\textbf{P2}^{\ast}$ when being constrained by the energy-limitation constraint $(\lambda_{B} < \mu_{B})$. Note that, due to the newly imposed energy-limitation constraint to $\textbf{P2}^{\ast}$, the feasible values of $\mu_{p}$ over which the linear program runs are given by
\begin{align}\label{Mp range for P2 with ELC}
\text{max} (\dfrac{\lambda_{p}}{1 - \delta}, f_{pd}) \leq \mu_{p} \leq f_{pd} + \left(1-f_{pd}\right) f_{ps}.
\end{align}

\begin{figure}[t!]
\centering
\includegraphics[width=0.8\columnwidth]{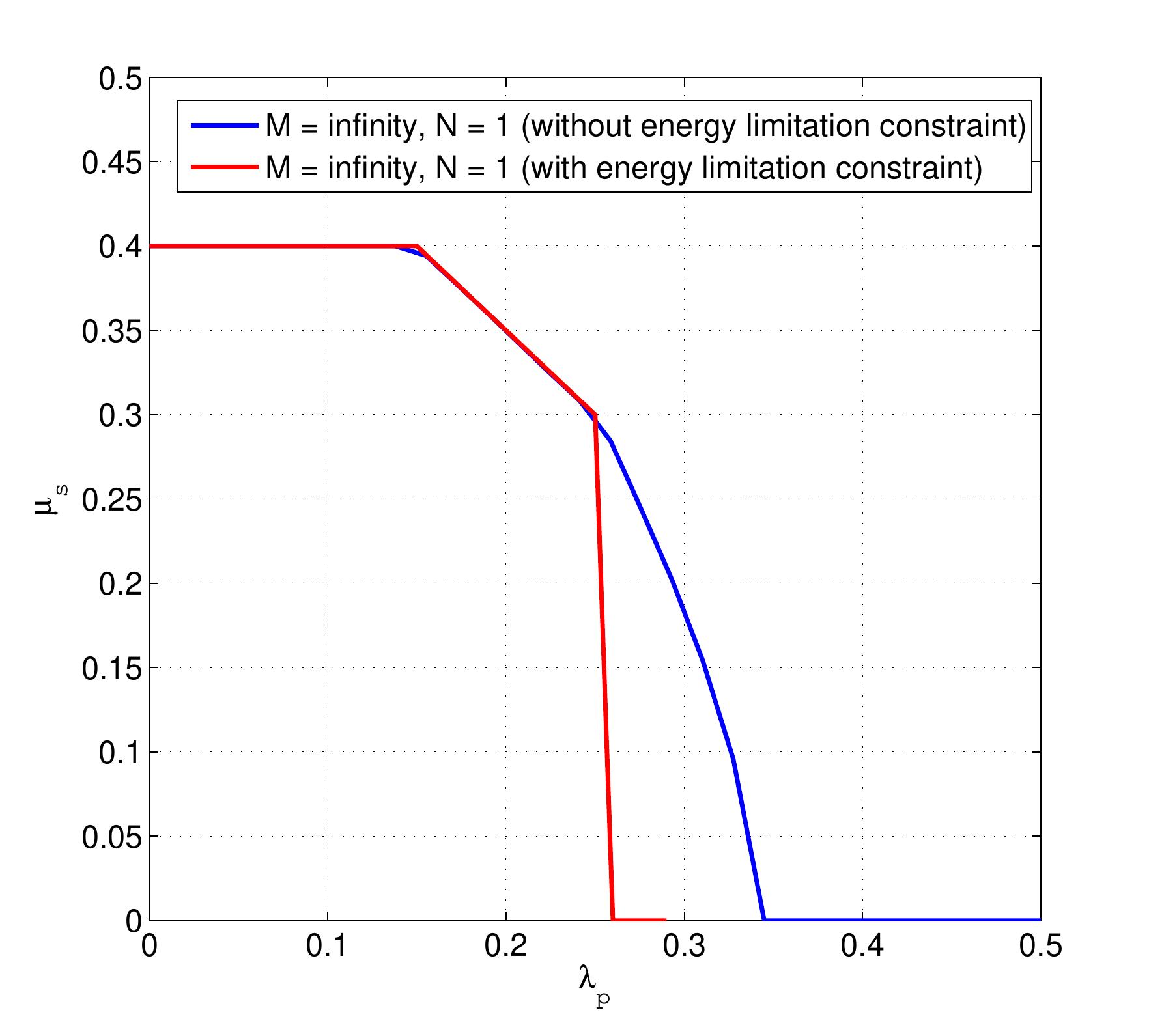}
    \caption{Comparing the stable throughout region of dominant system $2$ with its achievable one when imposing the energy limitation constraint.}
     \label{fig:13}
\end{figure}

 In Fig. \ref{fig:13}, we plot the achievable stable throughput region for dominant system $2$ with and without the energy limitation constraint \cite{Conference}. It is observed that the range of $\lambda_{p}$'s values, over which there exists a non-zero probability of having an empty $Q_{B}$ for dominant system $2$ without the energy limitation constraint, is $\lambda_{p} \leq 0.25$, where both systems achieve the same maximum SU's throughput. This is intuitive since for large values of $\lambda_{p}$, the steady state probability that $Q_{p}$ is empty and the SU is able to access the channel becomes very low that the number of energy packets inside $Q_{B}$ will grow to infinity, and $Q_B$ will always be capable of supporting the transmission of the SU's packets whenever it has the chance to access the channel.
\section{Numerical results}
\label{sec:num}
 In this section, our prime objective is to quantify the performance loss experienced by cooperative cognitive radio networks due to the existence of limited energy sources and finite queues, when compared to a baseline network with similar setup, yet, having an unlimited energy source, introduced in \cite{4,5}. Motivated by the sheer computational complexity of Algorithm \ref{euclid} for large values of queue lengths, as demonstrated in Section~\ref{sec:first}, we propose a heuristic scheme to characterize the stable throughput region of CCRNs with both finite battery and relay queues, and relatively large queue lengths. In the proposed heuristic scheme, we assume that all states have equal admission and selection probabilities, except for those having either deterministic admission or selection decision variables. Recall that we have demonstrated that the proposed heuristic scheme can be considered as a strong candidate to achieve a near-optimal stable throughput region for large queue lengths in Fig. \ref{fig:3}. Owing to the convexity of the proposed linear programs, we use standard optimization tools, e.g., CVX \cite{10}, to obtain the optimal solution. If not otherwise stated, we use the following parameters for our numerical results: $f_{pd} = 0.3$, $f_{ps} = 0.4$, $f_{sd} = 0.8$ and $\delta = 0.5$.
\begin{figure}[t!]
\centering
\includegraphics[width=0.8\columnwidth]{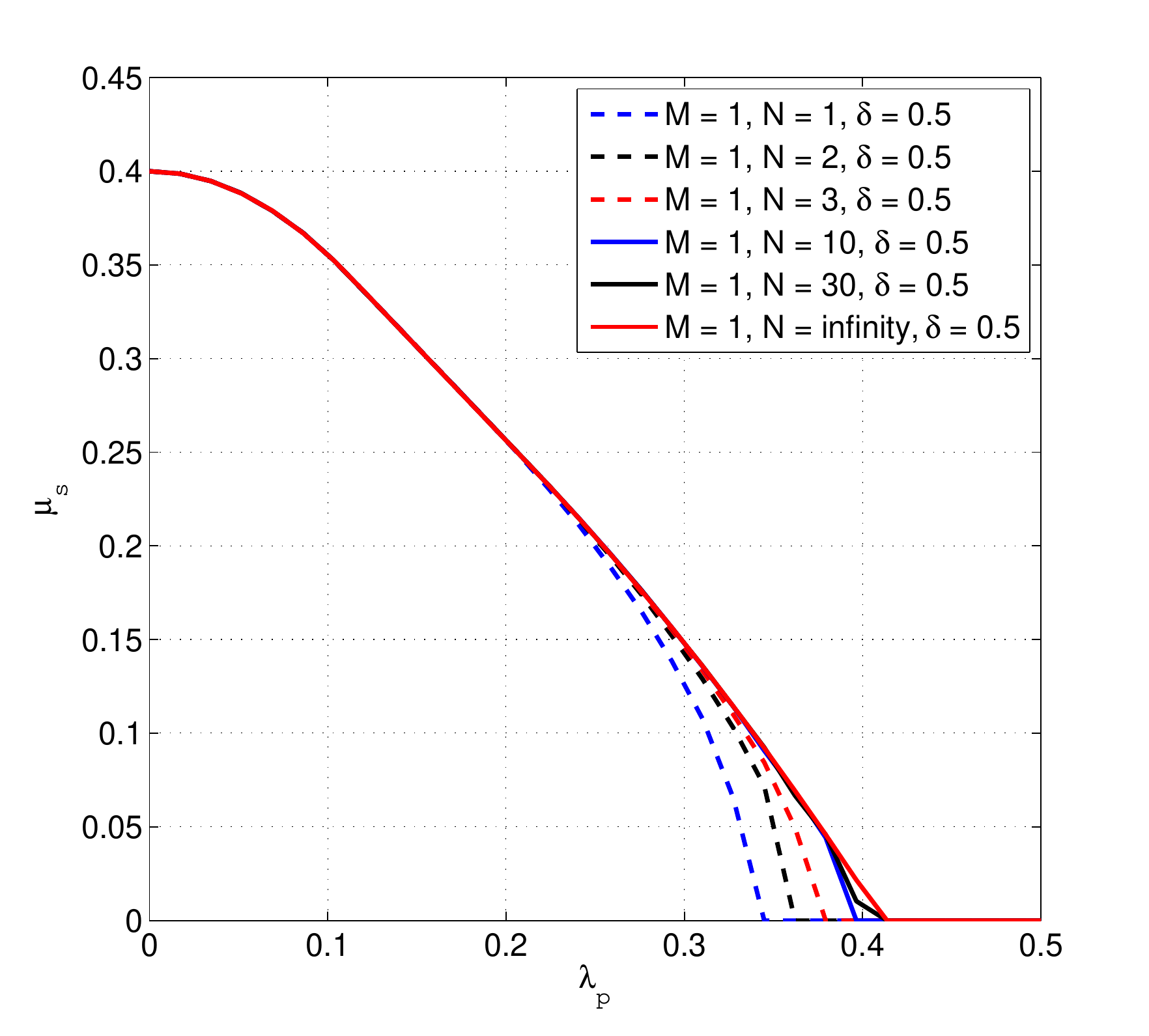}
    \caption{The achievable stable throughput region for different values of $N$.}
     \label{fig:14}
\end{figure} 

 Fig. \ref{fig:14} shows the effect of increasing $N$ on the achievable stable throughput region of CCRNs with both finite battery and relay queues. Towards that, we fix $M$ to $1$ and plot the stable throughput region for different values of $N$ $(N =$ $1$, $2$, $3$, $10$ and $30$). As expected, we observe that the maximum sustainable arrival rate of PU's data packets, which corresponds to a non-zero $\mu_{s}$, increases with $N$ due to increasing the number of available time slots for the SU to access the channel. We further observe that when $N = 30$, the achievable throughput region nearly approaches the one of dominant system $1$, introduced in Section \ref{sec:second}. This leads to an interesting insight that Generalized CCRNs with both finite battery and relay queues ($M = 1$ and $N = 30$) achieve the same stable throughput region of the system with infinite relay queue ($M = 1$ and $N =$ infinity). It is worth noting that, similar to Fig. \ref{fig:11}, $\lambda_{p} = 0.25$ divides the stable throughput region into two different regimes. However, the maximum SU's achievable throughput decreases from 0.3 to 0.1 due to reducing the length of the battery queue from infinity to 1.
\begin{figure}[t!]
\centering
\includegraphics[width=0.8\columnwidth]{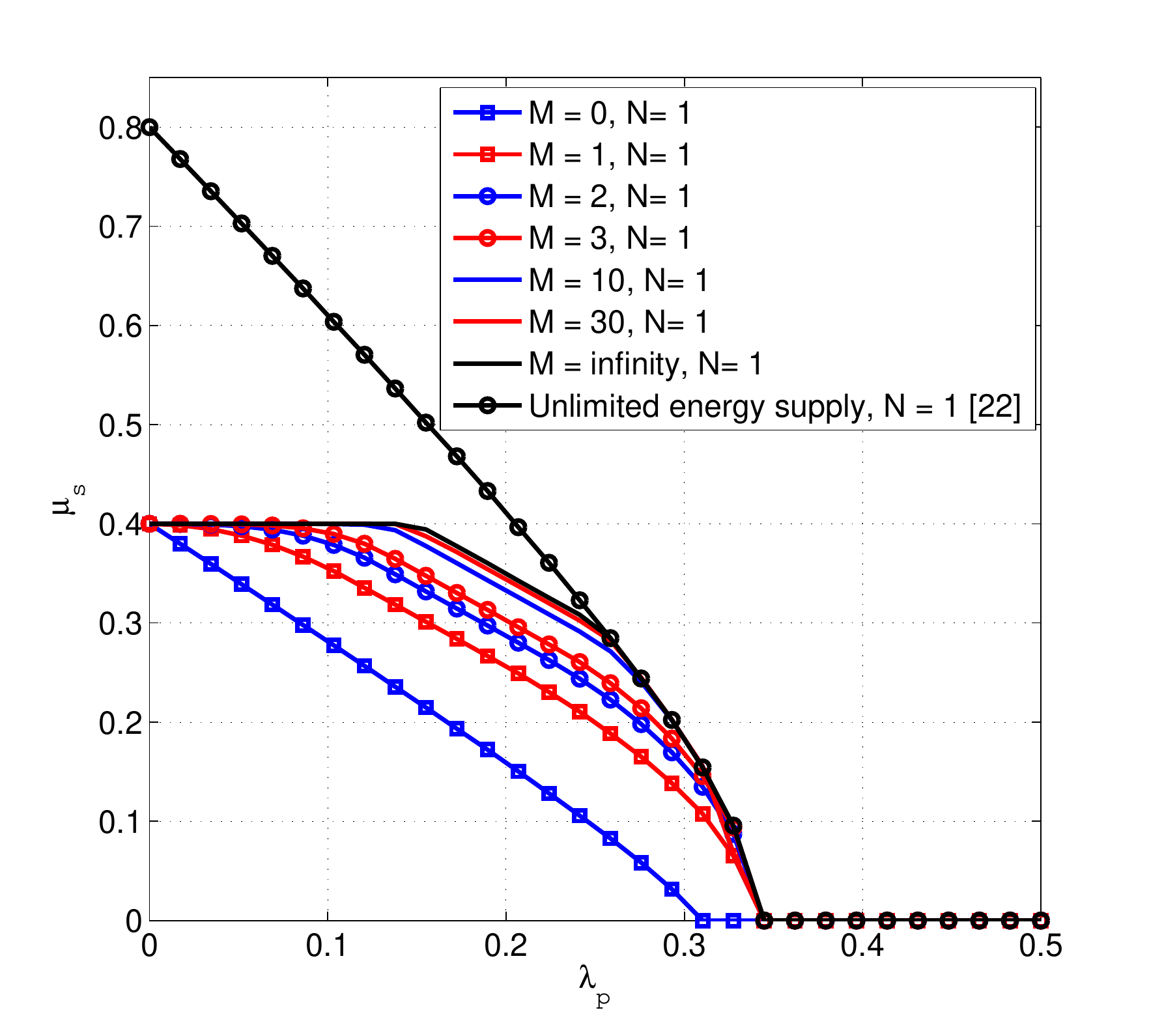}
    \caption{The achievable stable throughput region for different values of $M$.}
     \label{fig:15}
\end{figure}
 
In Fig. \ref{fig:15}, our objective is to demonstrate the impact of increasing $M$ on the achievable stable throughput region of CCRNs with both finite battery and relay queues. Towards this objective, we fix $N$ to $1$ and plot the stable throughput region for different values of $M$ $(M =$ $0$, $1$, $2$, $3$, $10$ and $30$). It is observed that the stable throughput region expands with increasing $M$ till it nearly approaches the same stable throughput region of the system with infinite battery queue ($M =$ infinity and $N = 1$), presented in Section \ref{sec:third}, when $M = 30$. It is further observed that the stable throughput region achieved by CCRNs with unlimited energy source and finite relay queue \cite{5} constitutes an upper bound on the achievable ones corresponding to different values of $M$. This is intuitive since the SU does not suffer from ''energy hunger" whenever it has the opportunity to access the channel in CCRNs introduced in \cite{5}.

\begin{figure}
\centering
\includegraphics[width=0.8\columnwidth]{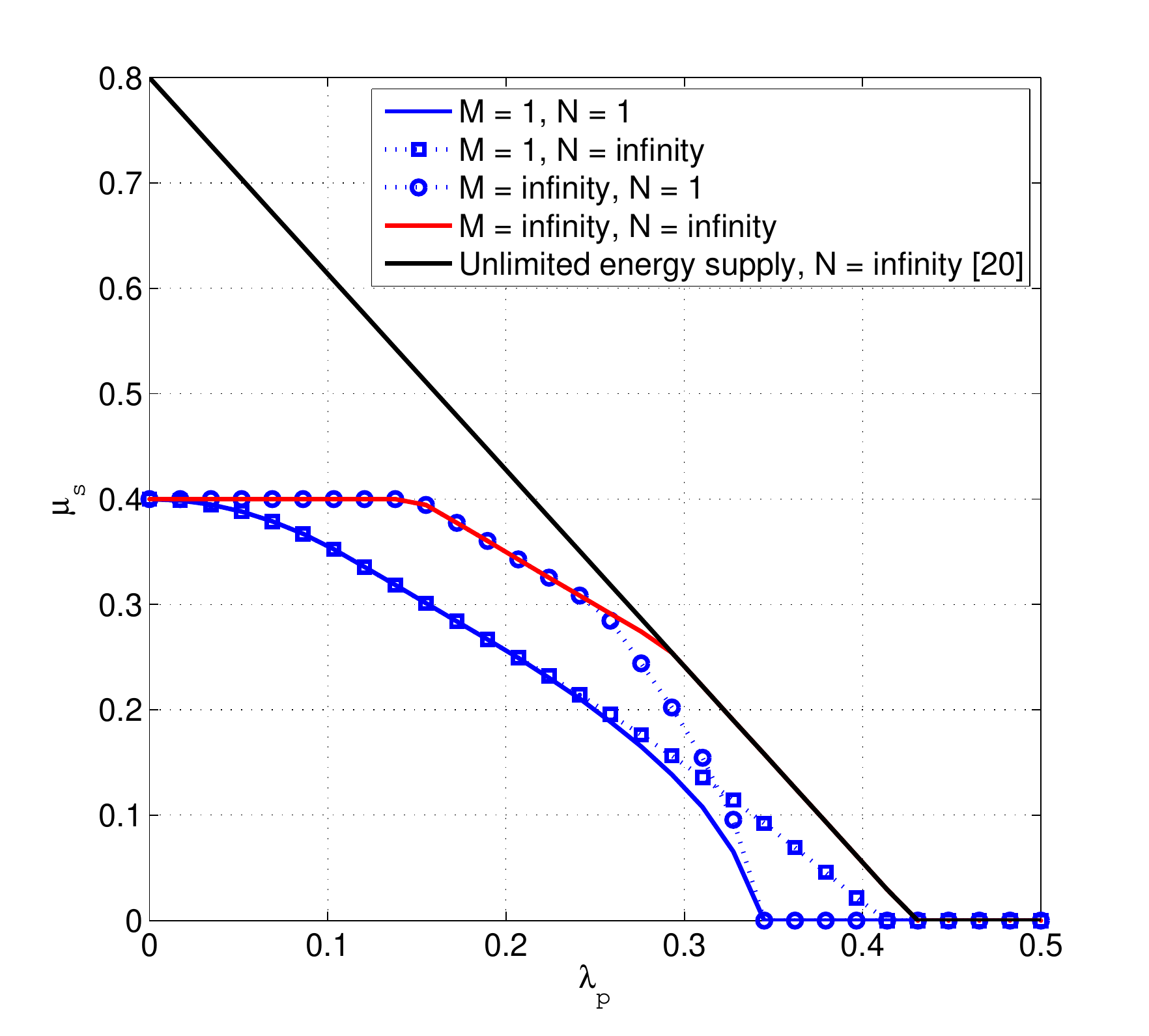}
    \caption{Comparing the achievable throughput regions by our proposed models with the of the baseline system with unlimited energy supply and infinite relay queue.}
     \label{fig:17}
\end{figure}

 Finally, Fig. \ref{fig:17} compares the achievable stable throughput regions of all studied CCRNs in this paper with that of the baseline system with infinite relay length and unlimited energy source \cite{4}. As expected, we observe that the achievable stable throughput region by \cite{4} constitutes an upper bound on the other systems (with finite queues and limited energy sources) due to the fact that the SU has no energy limitations and is equipped with infinite relay length at the same time. We further observe that, although the SU always has the chance to access the channel when $\lambda_{p} = 0$, the maximum SU's achievable throughput by all systems (including the system with both infinite battery and relay queues) is half the one achieved in \cite{4}. This, in turn, highlights the impact of the probabilistic arrival of energy packets at the SU $(\delta = 0.5)$.
\section{Conclusion}\label{sec:con}
In this paper, we studied a queuing-theoretic model for cooperative cognitive radio networks where the secondary user has a finite relay queue as well as a finite battery queue. We first developed an algorithm to characterize the stable throughput region numerically due to mathematical intractability towards obtaining closed-form expressions for the steady state distribution of the two-dimensional Markov Chain. Faced with this hurdle, we relaxed the system model and studied two simpler problems: 1) finite battery queue with infinite relay queue and 2) finite relay queue with infinite battery queue. The stable throughput regions were characterized for the simpler systems. Specifically, we formulated the stable throughput region optimization problem for problem and showed how to solve it. Finally, we compared the achievable throughput region of all studied systems with that of the baseline system with unlimited energy sources and infinite queues. Our numerical results quantified the expansion in the throughput region due to increasing the battery queue size, and the enhancement of the maximum sustainable arrival rate of PU's data packets, corresponding to a non-zero SU's achievable throughput, due to increasing the relay queue size. Furthermore, they revealed that finite battery and relay queues of sufficiently large sizes are enough to achieve the same level of benefits of a system with infinite queue sizes. They also showed the profound role of the arrival rate of the energy harvesting process at the SU on the achievable stable throughput region.

 This work has many possible extensions. For instance, we focused in this paper only on characterizing the stable throughput region of cooperative cognitive radio networks with both finite battery and relay queues. One possible extension is to study the delay analysis of our finite queue lengths cooperative cognitive radio networks. In addition, another possible extension is to investigate the achievable stable throughput region when the arrival rate of the energy harvesting process at the SU may not be enough to keep the node alive to listen to the channel at every slot. In such scenario, even when using the Dominant System approach, the service rate of the battery queue at the SU will not only depend on the PU's state. This, in turn, complicates the associated two-dimensional Markov Chain and, hence, leads to a significantly higher computational complexity of stable throughput region characterization, which possibly calls for new models and approaches. Another possible future work is to consider the scenario of having multiple secondary users.


\bibliographystyle{IEEEtran}
\bibliography{Term_paper}

\end{document}